\documentclass[fleqn,usenatbib]{mnras}

\usepackage{newtxtext,newtxmath}

\usepackage[T1]{fontenc}

\DeclareRobustCommand{\VAN}[3]{#2}
\let\VANthebibliography\thebibliography
\def\thebibliography{\DeclareRobustCommand{\VAN}[3]{##3}\VANthebibliography}

\usepackage{graphicx}	
\usepackage{amsmath}	
\usepackage{amssymb}	

\title[dEB SB2 J064726.39+223431.6]{Fundamental parameters for double-lined spectroscopic and detached eclipsing binary system J064726.39+223431.6.}

\author[M. Kovalev et al.]{
Mikhail Kovalev,$^{1,2,3,4}$\thanks{E-mail: mikhail.kovalev@ynao.ac.cn}
Song {Wang},$^{5}$
Xuefei Chen$^{1,2,6}$
and Zhanwen Han$^{1,2,6}$
\\
$^{1}$Yunnan Observatories, China Academy of Sciences, Kunming 650216, China\\
$^{2}$Key Laboratory for the Structure and Evolution of Celestial Objects, Chinese Academy of Sciences, Kunming 650011, China\\
$^{3}$Sternberg Astronomical Institute, M. V. Lomonosov Moscow State University, Leninskie Gory, Moscow 119991, Russia\\
$^{4}$Key Laboratory of Optical Astronomy, National Astronomical Observatories, Chinese Academy of Sciences, Beijing 100101, China\\
$^{5}$Center for Astronomical Mega-Science, Chinese Academy of Sciences, 20A Datun Road, Chaoyang District, Beijing 100012, China\\
}

\date{Accepted XXX. Received YYY; in original form ZZZ}

\pubyear{2022}

\def\kms{\,{\rm km}\,{\rm s}^{-1}}

\def\feh{\hbox{[Fe/H]}}

\newcommand{\teff}{T_{\rm eff}}
\newcommand{\rv}{{\rm RV}}
\def\Vmic{V_{\rm mic}}

\def\vsini{V \sin{i}}
\def\logg{\log{\rm (g)}}
\def\snr{\hbox{S/N}}
\newcommand{\ha}{\hbox{H$\alpha$}}

\begin{document}
\label{firstpage}
\pagerange{\pageref{firstpage}--\pageref{lastpage}}
\maketitle

\begin{abstract}
We present a study of the detached eclipsing binary J064726.39+223431.6 using spectra from the LAMOST medium-resolution spectra and TESS photometry. We use full-spectrum fitting to derive radial velocities and spectral parameters: ${\teff}_{A,B}=6177,\,5820$ K, $\vsini_{A,B}=59,\,50~\kms$ and $\feh_{A,B}=-0.19$ dex. The orbital solution and light curve analysis suggest that it is a close pair of fast rotating stars on a circular orbit. We measure their masses to be $M_{A,B}=1.307\pm0.007,\, 1.129\pm0.005\,M_\odot$ and their radii to be $R_{A,B}=1.405\pm0.052,\, 1.219\pm0.060\,R_\odot$ resulting in surface gravities of $\logg_{A,B}=4.259\pm0.033,\,4.319\pm0.042$ (cgs). Theoretical models cannot match all of these properties, predicting significantly higher $\teff$ for a given metallicity. The derived age of the system is 1.56 Gyr, which indicates that both components are younger than Sun, which contradicts to much longer orbit's circularisation timescale.

\end{abstract}




\section{Introduction}

Binary stellar systems are very important objects for astronomy, since they allow us to learn more than from single stars. For example, observations of the periodic eclipses and changes in line-of-sight velocities (RV) can be used to directly measure sizes and masses of the stellar components, if orbital inclination is high enough. 
 Such dynamical measurements in detached eclipsing binaries (dEBs) minimally rely on stellar models \citep{1991A&ARv...3...91A,torres2010}, compared to other methods such as ``gravity mass" (estimated from parallax, extinction, magnitude, effective temperature, and surface gravity) or ``evolutionary mass" (derived from theoretical isochrones with the effective temperature, surface gravity, and metallicity as input).

 The well-known catalogue DEBCat includes more than 300 dEB systems with dynamical mass and radius measurements to the 2\% precision \citep{2015ASPC..496..164S}.
Recently, by using the LAMOST (Large Sky Area Multi-Object fiber Spectroscopic Telescope, also known as Guoshoujing telescope) medium-resolution spectral (MRS) data and other time-domain photometric data, \cite{xiong} derived independent atmospheric parameters and accurate masses and radii for 56 new dEBs. These measurements with high precision can be used to study the correlations of different physical parameters (e.g., the mass-luminosity, mass-radius, and mass-effective temperature relations) and provide useful constraints on theoretical models of stellar evolution \citep{torres2010,2015AJ....149..131E,2018MNRAS.479.5491E,zw2020}.

More than 2400 double-lined spectroscopic binaries (SB2) were identified in \cite{cat22} based on the LAMOST-MRS survey \citep{mrs} and we selected one previously known dEB star ZTF J064726.39+223431.6 with large RV semiamplitude, based on \cite{songK2}, to estimate physical parameters for both components in this system. This SB2 is also contained by the new ``High-precision Empirical Stellar Mass Library" constructed by \cite{xiong}. Table~\ref{tab:des} list the designations of this system in literature. 

In this paper we use the LAMOST-MRS data and additional photometric data to measure physical parameters for both stars in this system and test our approach of multiple-epoch SB2 fitting, developed in \cite{tyc}.
The paper is organised as follows: in Sections~\ref{sec:obs} and \ref{sec:methods} we describe the observations and methods. Section~\ref{results} presents our results. In Section~\ref{discus} we discuss the results in context of binary system evolution. In Section~\ref{concl} we summarise the paper and draw conclusions.

\begin{table}
    \centering
    \caption{Designations from the literature: a-\protect\cite{gaia3}, b-\protect\citet{varstarindex}, c-\protect\cite{tic}, d-\protect\cite{asassnv}, e-\protect\cite{ztf_var}, f-\protect\cite{qian19}.}
    \begin{tabular}{lcc}
    \hline
    Property & Value & Reference\\
    \hline
    Gaia DR3  &  3378682653860701568  & a \\
    Variable Star indeX &  167003  & b \\
    TESS input catalogue & TIC57046871 & c\\
    ASAS-SN & J064726.41+223431.7 & d\\
    ZTF & J064726.39+223431.6 & e\\
    LAMOST & J064726.39+223431.7 & f\\
    \hline
    \end{tabular}
    \label{tab:des}
\end{table}

\section{Observations}
\label{sec:obs}
\subsection{Spectra}

LAMOST is a 4-meter quasi-meridian reflective Schmidt telescope with 4000 fibers installed on its 5-degree-FoV focal plane. These configurations allow it to observe spectra for at most 4000 celestial objects simultaneously (\cite{2012RAA....12.1197C, 2012RAA....12..723Z}).
 All available spectra were downloaded from \url{www.lamost.org} under the designation J064726.39+223431.7.	We use the spectra taken at a resolving power of R$=\lambda/ \Delta \lambda \sim 7\,500$. Each spectrum is divided into two arms: blue from 4950\,\AA~to 5350\,\AA~and red from 6300\,\AA~to 6800\,\AA. We convert the heliocentric wavelength scale in the observed spectra from vacuum to air using \texttt{PyAstronomy} \citep{pya}. Observations are carried out from 2019-11-19 till 2021-02-19, covering 16 nights with time base of 480 days.
 Since the period is short ($P=1.217$ d$\sim29$ hrs) we analysed spectra taken during short 20 minutes exposures, unlike \cite{tyc} and \cite{cat22}, where spectra stacked for the whole night were used. We discarded all spectra taken on three nights with MJD=58829,~59216,~59249 d due to excess noise. In total we have 72 spectra, where the average signal-to-noise ratio ($\snr$) of a spectrum ranges from 16 to 54 ${\rm pix}^{-1}$  and the majority of the spectra have $\snr$ around 40 ${\rm pix}^{-1}$.

\subsection{Photometry}

We have downloaded publicly available ZTF light curves (LC)\footnote{\url{https://irsa.ipac.caltech.edu/cgi-bin/ZTF/nph_light_curves}}. These LCs contain 1444 datapoints in $r$ band and 370 datapoints in $g$ bands and cover timebase 1240 d and 1125 d respectively. We also download LC in $V$ band from the ASSAS-SN portal\footnote{\url{https://asas-sn.osu.edu/variables/AP24824756}}. It contains only 183 datapoints and covers timebase 1212 d. After phase-folding of these LCs with the period we found that only ZTF LC from the $r$ band has good coverage of eclipses, therefore we use only it in the further analysis.    
\par
\begin{figure*}
    \includegraphics[width=\textwidth]{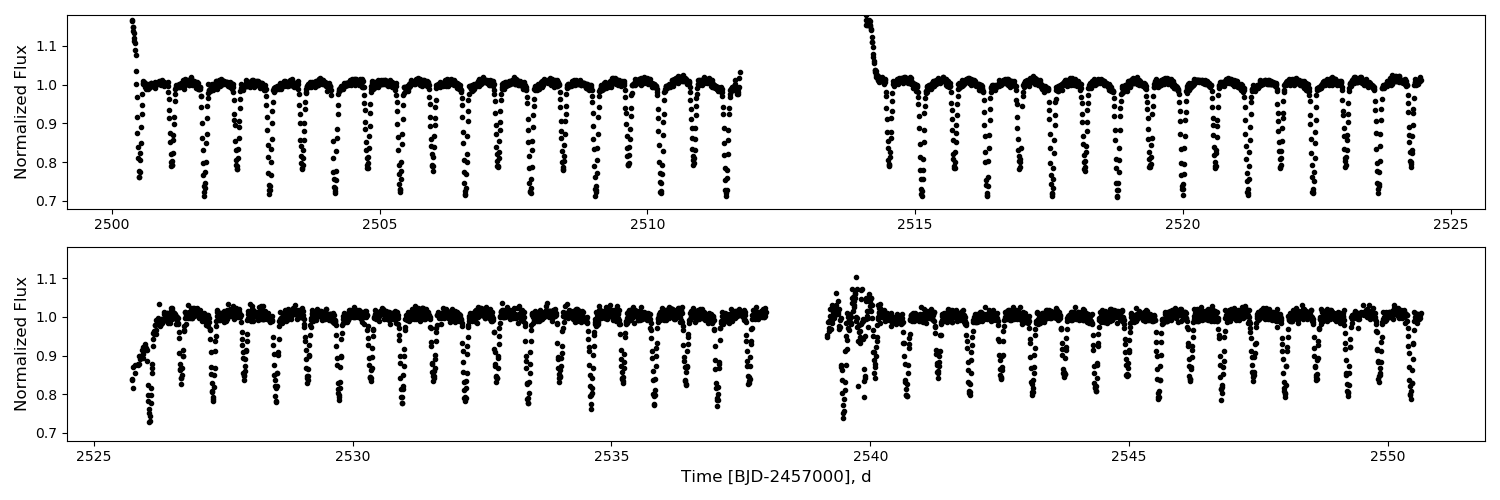}
    \caption{Light curves extracted by \texttt{eleanor} from TESS sectors 44 (top) and 45 (bottom).}
    \label{fig:tess2}
\end{figure*}

The Transiting Exoplanet Survey Satellite \citep[TESS][]{tess} mission observed this star in two sectors 44 and 45, which covers $\sim27$ days each. The LCs are not available on the MAST\footnote{\url{https://mast.stsci.edu/portal/Mashup/Clients/Mast/Portal.html}} portal yet, therefore we use \texttt{eleanor} \citep{eleanor,astrocut} to extract the LC datasets. We use default settings and clip the edges of the LC, as they have some processing artifacts, see Figure~\ref{fig:tess2}.   
After the clipping, LCs contain 2953 and 3160 datapoints for sectors 44 and 45, respectively, and have relative (non-calibrated) stellar magnitudes. The background subtraction is not optimal for these datasets, therefore we analyse them separately.

\section{Methods} 
\label{sec:methods}

\subsection{Spectral fitting}
\label{sec:maths} 

Our spectroscopic analysis includes two consecutive stages: 
\begin{enumerate}
    \item analysis of individual observations by binary and single-star spectral models, where we normalise the spectra and make a rough estimation of the spectral parameters, see brief description in Section~\ref{sec:ind}. 
    \item simultaneous fitting of multiple-epochs with a binary spectral model, using constraints from binary dynamics and values from the previous stage as an input, see Section~\ref{sec:multi}.
\end{enumerate}
A LAMOST-MRS implementation of this method was first presented in \citet{tyc}. 

\subsubsection{Individual spectra.}
\label{sec:ind}
The single-star spectral model is described in Appendix~\ref{sec:payne}.
The normalised binary model spectrum is generated as a sum of the two Doppler-shifted, normalised single-star model spectra ${f}_{\lambda,i}$ scaled according to the difference in luminosity, which is a function of the $\teff$ and stellar size. We assume both components to be spherical and use following equation:    

\begin{align}
    {f}_{\lambda,{\rm binary}}=\frac{{f}_{\lambda,2} + k_\lambda {f}_{\lambda,1}}{1+k_\lambda},~
    k_\lambda= \frac{B_\lambda(\teff{_{,1}})~M_1}{B_\lambda(\teff{_{,2}})~M_2} 10^{\logg_2-\logg_1}
	\label{eq:bolzmann}
\end{align}
 where  $k_\lambda$ is the luminosity ratio per wavelength unit, $B_\lambda$ is the black-body radiation  (Plank function), $\teff$ is the effective temperature, $\logg$ is the surface gravity and $M$ is the mass. Throughout the paper we always assume the primary star to be brighter.
\par
The binary model spectrum is later multiplied by the normalisation function, which is a linear combination of the first four Chebyshev polynomials \citep[similar to][]{Kovalev19}, defined separately for blue and red arms of the spectrum. The resulting spectrum is compared with the observed one using the \texttt{scipy.optimise.curve\_fit} function, which provides optimal spectral parameters, radial velocities (RV) of each component plus the mass ratio and two sets of four coefficients of Chebyshev polynomials. We keep metallicity equal for both components. In total we have 18 free parameters for a binary fit. We estimate goodness of the fit parameter by reduced $\chi^2$:

\begin{flalign}
\label{eq:chi2}
   \chi^2 =\frac{1}{N-18} \sum \left[ \left({f}_{\lambda,{\rm observed}}-{f}_{\lambda,{\rm model}}\right)/{\sigma}_{\lambda}\right]^2
\end{flalign}
where $N$ is a number of wavelength points in the observed spectrum. To explore the whole parameter space and to avoid local minima we run the optimisation six times with different initial parameters of the optimiser. We select the solution with minimal  $\chi^2$ as a final result. 
\par 
Additionally, every spectrum is analysed by a single star model, which is identical to a binary model when both components have all equal parameters, so we fit only for 13 free parameters. Using this single star solution we compute the difference in reduced $\chi^2$ between two solutions and the improvement factor, computed using Equation~\ref{eqn:f_imp} similar to \cite{bardy2018}. This improvement factor estimates the absolute value the difference between the two fits and weights it by difference between two solutions.

\begin{align}
\label{eqn:f_imp}
f_{{\rm imp}}=\frac{\sum\left[ \left(\left|{f}_{\lambda,{\rm single}}-{f}_{\lambda}\right|-\left|{f}_{\lambda,{\rm binary}}-{f}_{\lambda}\right|\right)/{\sigma}_{\lambda}\right] }{\sum\left[ \left|{f}_{\lambda,{\rm single}}-{f}_{\lambda,{\rm binary}}\right|/{\sigma}_{\lambda}\right] },
\end{align}
where ${f}_{\lambda}$ and ${\sigma}_{\lambda}$ are the observed flux and corresponding uncertainty, ${f}_{\lambda,{\rm single}}$ and ${f}_{\lambda,{\rm binary}}$ are the best-fit single-star and binary model spectra, and the sum is over all wavelength pixels.

\subsubsection{Multiple-epochs fitting.}
\label{sec:multi}

We explore the results from the fitting of the individual epochs and find that a result's quality clearly depends on separation of RVs. Clear double-lined spectra show that spectral lines are significantly broadened ($\vsini \sim 60~\kms$), thus at phases near the conjunctions our fitting algorithm often cannot reliably separate primary/secondary contributions. Fortunately we can automatically separate good results with clear double-lines using $f_{\rm imp}$. Thus we decided to use only five individual spectra with highest $f_{\rm imp}$ in multiple-epoch fitting. 
\par
If two components in our binary system are gravitationally bound, their radial velocities should agree with the following equation:
\begin{align}
\label{eqn:asgn}
    {\rm RV_{A}}=\gamma_{\rm dyn} (1+q_{\rm dyn}) - q_{\rm dyn} {\rm RV_{B}},
\end{align}
where $q_{\rm dyn}={M_B}/{M_{A}}$ - mass ratio of binary components and $\gamma_{\rm dyn}$ - systemic velocity. Using this equation we can directly measure the systemic velocity and mass ratio. 
\par
\par
To reduce the number of free parameters in the multiple-epochs fitting we use only ${\rm RV_A}$ and computed ${\rm RV_B}$ using Equation~\ref{eqn:asgn}. The same value of mass ratio is used in Equation~\ref{eq:bolzmann}. Unlike the previous stage, we fit $\feh$ for both components. In total we fit for 15 free parameters.   
We fit five previously normalised individual epoch's spectra with maximal improvement factor, using their binary spectral parameters values for initialisation. We select the solution with minimal  $\chi^2$ as a final result.

\subsubsection{Typical errors estimation}
\label{sec:err}
We estimate the typical errors of the multiple-epochs fitting by testing it's performance on the dataset of synthetic binaries. We generated 3000 mock binaries using uniformly distributed mass-ratios from 0.7 to 1.0, $\teff$ from $4600$ to $8400$ K, $\logg$ from 2.6 to 4.6 (cgs) and $\vsini$ from 1 to 100 $\kms$. Metallicity was set to $\feh=0.0$ dex in both components. For each star, we computed 5 mock binary spectra using radial velocities computed for circular orbits with the semiamplitude of the primary component $20\,\kms$ at randomly chosen phases. These models were degraded by Gaussian noise according to $\snr=100$ pix$^{-1}$. 
We performed exactly the same analysis on this simulated dataset as we did for the observations. We checked how well the mass ratio and the spectral parameters of the primary and secondary components can be recovered by calculating the average and standard deviation of the residuals. For the primary components we have $\Delta\teff=95\pm239$~K, $\Delta\logg=0.07\pm0.14$ cgs units, $\Delta\vsini=-1\pm12\kms$  and $\Delta\feh=-0.02\pm0.08$ dex. For the secondary components we have $\Delta\teff=35\pm364$~K, $\Delta\logg=0.05\pm0.21$ cgs units, $\Delta\vsini=-3\pm21\kms$ and $\Delta\feh=0.04\pm0.18$ dex. The mass ratio recovery has $\Delta q=0.05\pm0.15$.

\subsection{Orbital fitting}
To get the orbital solution, we select 55 RVs of the primary component separated by at least $60~\kms$ from the systemic velocity $\gamma$. RVs close to $\gamma$ can be affected by the Rossiter-McLaughlin (RM) effect \citep[][]{rossiter,mclaflin}, as several spectra were possibly taken during eclipses, thus we don't use them in orbital fitting. We collect all RV measurements in Table~\ref{tab:rvs} 
\par
In the next step, selected ${\rm RV_{A}}$ are used to fit circular orbits using generalised Lomb-Scargle periodogram (\texttt{GLS}) code by \cite{gls} :

\begin{align}
    {\rm RV}_A(t)=\gamma- K_A \sin \left (\frac{2\pi}{P}(t -t_0) \right ),
\end{align}
where $\gamma$ - is the systemic velocity, $P$ - is the period, $t_0$ -is the conjunction time, $K$ - is the radial velocity semiamplitude.
 We also fit to a Keplerian orbit and find that the eccentricity is equal to zero, so a circular orbit is a valid assumption.

\begin{table}
\begin{center}
\caption{{\sc GLS} orbital solution.}
\begin{tabular}{lc}
\hline
Parameter & Value\\
\hline
$P$, d & $1.217770 \pm 0.000003$\\
$t_0$, HJD d & $2458808.8088 \pm 0.0003$\\
$K_A,\,\kms$ & 122.80$\pm$0.23\\
$\gamma_A,\, \kms$ &$-16.98 \pm0.16$\\
\hline
\label{tab:orbit}
\end{tabular}
\end{center}
\end{table} 

\subsection{Light curve fitting}

We used the \texttt{JKTEBOP} code (version 40)\footnote{{\sc jktebop} is written in {\sc fortran77} and the source code is available at \url{http://www.astro.keele.ac.uk/jkt/codes/jktebop.html}} by \citet{jkt} to simultaneously fit the LC and RV timeseries. We used only ZTF $r$ band and TESS datasets, because of their better coverage of the eclipses. Unlike \texttt{GLS}, \texttt{JKTEBOP} allows us to fit RVs measured for both components. Our fitting was initialised using $P,t_0,\gamma$ values from the \texttt{GLS} fit. We used tables of the limb and gravity darkening coefficients from \citet{claret2011} and \cite{claret2017}, and linearly interpolated them for spectral parameters from multiple-epochs fits. We used four-parameter limb darkening coefficients $c_i,\,i=1,4$. We took into account a useful comment by \citet{torres21} on the usage of gravity darkening coefficients in \texttt{JKTEBOP}. 

The systemic velocity was fitted for both binary components. Additionally, we fit for a ``third" light contribution $L_3$ and the nuisance parameter the out-of-eclipse magnitude $S_0$ for all three datasets. In total we fit for 16 parameters: $J$ the central surface brightness ratio, $(R_A+R_B)/a$ the ratio of the sum of stellar radii to the semimajor axis, $R_B/R_A$ the ratio of the radii, $i$ the inclination, $e\cos{\omega}$, $e\sin{\omega}$  the eccentricity multiplied by the cosine and sine of the periastron longitude, ${\rm reflected~light}_{A,B}$, $P$ the period, $t_0$, semiamplitudes and systemic velocities $K_{A,B}$, $\gamma_{A,B}$, $L_3$ and $S_0$. We use integration ring size $5^{\circ}$ for the ZTF $r$ and $1^{\circ}$ for the TESS LCs.  

At first we run the \texttt{JKTEBOP} code in the mode ``Task 4'' to discard outliers larger than three sigma and allow it adjusting observational errors through several iterations until reduced $\chi^2$ reaches unity. This removes 24, 17, 13 datapoints from ZTF $r$, TESS 44, 45 LC datasets, respectively and keeps all 110 RV datapoints. Then we run \texttt{JKTEBOP} in the residual permutation (RP) mode (Task 9 - "red" noise) to estimate uncertainties using 1417, 2935 and 3147 residual-shifted fits for ZTF $r$ and TESS 44, 45 datasets respectively, similarly to \citet{southworth2011,jkt}. 

\begin{figure*}
	\includegraphics[width=\textwidth]{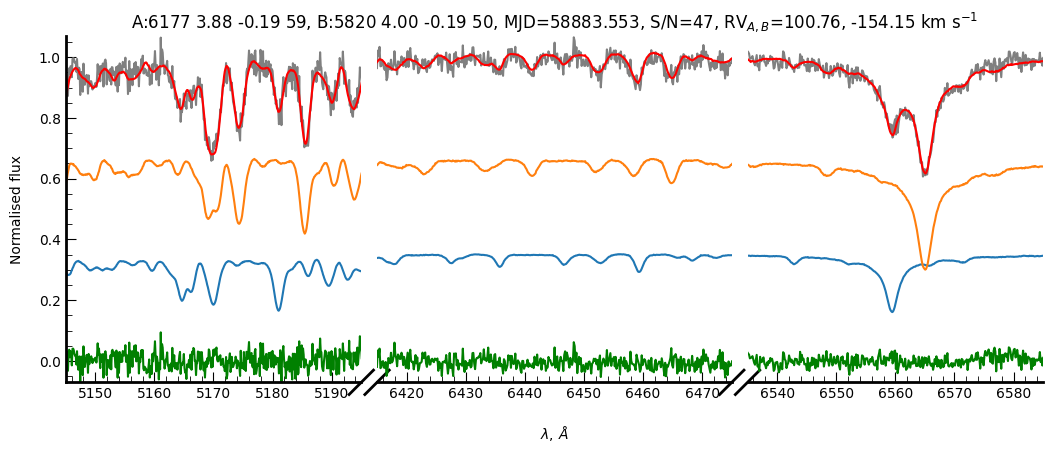}
	\includegraphics[width=\textwidth]{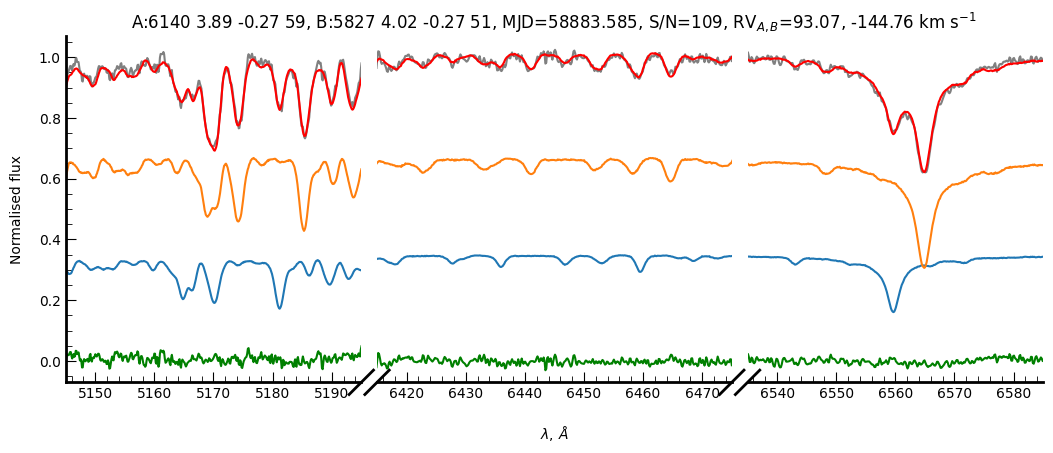}
    \caption{Example of the multiple-epoch fitting for short epochs spectra (top) and stacked spectra for whole night (bottom). The observed spectrum is shown as a gray line, the best fit is shown as red line. The primary component is shown as the orange line, the secondary as a blue line. The fit residuals are shown as a green line.}
    \label{fig:spfit}
\end{figure*}

\section{Results}
\label{results}
In the Figure~\ref{fig:spfit} we show the best fit by multiple-epochs binary model for an epoch with large RV separation. We zoom into the wavelength range around the magnesium triplet and $\ha$ and in a 70~\AA~interval in the red arm, where many double lines are clearly visible. Both components are on main sequence, where the primary star ($\teff=6104~K, \logg=3.86$ cgs, $\feh=-0.29$ dex, $\vsini=58~\kms$) contributes around 70\% in the visible light, while the secondary star ($\teff=5980~K, \logg=4.07$ cgs, $\feh=0.02$ dex, $\vsini=50~\kms$) and contributes to the remaining 30\%. Primary RVs derived in multiple-epoch fit are almost identical to ones estimated from fits for individual spectra. The derived mass ratio is $q\sim0.87$ and systemic radial velocity $\gamma=-17.63~\kms$. Additionally we make another multiple-epoch fit assuming same metallicity for both components, finding a slight change in $\teff$ and $\logg$ with two components being metal-poor with $\feh=-0.19$ dex. We can use $q$ and $\Delta\logg$ to estimate ratio of stellar sizes $R_B/R_A=0.73, 0.81$ for a solution with different and same metallicity. Additionally we fitted five out of 16 stacked spectra, previously analysed in \cite{cat22}. These spectra have significantly higher $\snr$ (see bottom panel of Figure~\ref{fig:spfit}), but their lines are slightly blurred due to changes in $\rv$. As a result we derived smaller metallicities ($\Delta\feh\sim-0.07$ dex in comparison with short epochs results), although other parameters were almost identical to previous estimates for short epoch's spectra. We present all spectroscopic results in the Table~\ref{tab:final}. The errors in the spectral parameters provided by \texttt{scipy.optimise.curve\_fit} are nominal and largely underestimated. We also compared our spectral solutions for SB2 components with spectra, separated using other disentangling methods in Appendix~\ref{separation} and found a good agreement.
\par
\begin{figure}
	\includegraphics[width=\columnwidth]{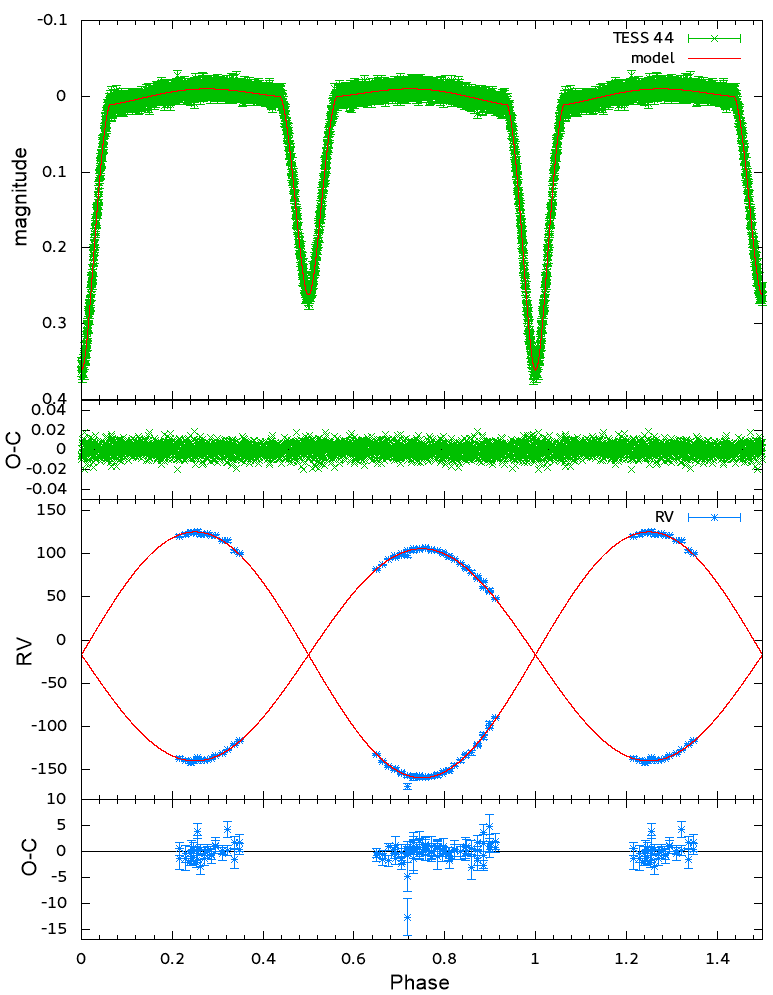}
    \caption{Phase-folded LC from TESS44 (top) and orbit (bottom) fits with \texttt{JKTEBOP}. In the bottom panels we show fit residuals O-C. The magnitudes are not calibrated.}
    \label{fig:orbit_tess44}
\end{figure}

 
In Figure~\ref{fig:orbit_tess44} we show the best fits of the TESS 44 LCs (top) and RVs (bottom) by \texttt{JKTEBOP}.  Fits for the ZTF~$r$ and TESS~45 can be found in Figures \ref{fig:orbit_rg},\ref{fig:orbit_tess45}. The fit residuals (O-C) are typically small ($\leq 0.10$ mag for ZTF~$r$, $\leq 0.02$ mag for TESS and $\leq 5~\kms$ for RVs). Only one measurement for $\rv_B$ is off by $\sim-15~\kms$. The derived systemic velocities are very similar to the values derived from a multiple-epoch spectral fit. Orbital solutions for the primary star from \texttt{GLS} and \texttt{JKTEBOP} agree well, taking to account uncertainties. A very small value for the eccentricity ($e\leq0.002$) confirms that orbit is  circular or very close to it. The orbital solution from Table~3 in \cite{songK2} has $e=1$ and $K_A=544\,\kms$, which is wrong due to failed optimisation.
Third light contributions are significant in both TESS datasets (12 and 29 per cent) and negligible in ZTF $r$. This is not surprising due to the small aperture of TESS cameras relative to one in telescopes used for ZTF.   
The oblateness of the components ($0.017, 0.007$ for ZTF $r$ and $0.013,0.011$ for TESS) is small ($\leq0.04$), thus \texttt{JKTEBOP} provides a reliable solution for this system \citep[][]{popper_etzel}. We show ``corner" plots with all RP simulations in Figures~\ref{fig:corner},\ref{fig:corner2},\ref{fig:corner3}, where correlations between fitted parameters can be explored.
We present results from \texttt{GLS} and \texttt{JKTEBOP} in the Tables~\ref{tab:orbit},\ref{tab:lcorbit}\footnote{ Two additional LC datasets (ZTF $g$ and ASAS-SN $V$) were also fitted, but provided poor results inconsistent with ZTF $r$, TESS and spectroscopic fits ($R_B>R_A$), possibly due to poor coverage of eclipses. }

\begin{table*}
    \centering
        \caption{{\sc JKTEBOP} solutions for ZTF $r$ and TESS LCs. Error's estimates are from RP simulations.}
    \begin{tabular}{l|ccc}
\hline
Parameter & ZTF $r$ & TESS 44 & TESS 45\\
\hline
fixed:\\
Grav. darkening$_A$ & 0.3108& \multicolumn{2}{c}{0.2576}\\
Grav. darkening$_B$ & 0.3317& \multicolumn{2}{c}{0.2760}\\
Limb darkening $c_{A,i}$ & 0.3029, 0.8993, -0.7099, 0.2098& \multicolumn{2}{c}{0.4002, 0.4901, -0.3288, 0.0638}\\
Limb darkening $c_{B,i}$ & 0.3420, 0.7173, -0.4286, 0.0838& \multicolumn{2}{c}{0.4427, 0.3238, -0.0893, -0.0390}\\
\hline
fitted:\\
$J$  &0.762$\pm$0.025 & 0.770$\pm$0.005& 0.772$\pm$0.019 \\
$(R_A+R_B)/a$  &0.3935$\pm$0.0065 & 0.4063$\pm$0.0013 &0.4052$\pm$0.0033 \\
$R_B/R_A$  &0.667$\pm$0.065 & 0.868$\pm$0.076 &0.840$\pm$0.152 \\
$i^{\circ}$  & 83.77$\pm$1.74 &81.66$\pm$0.29 &81.13$\pm$0.74 \\
$e\cos{\omega}$  &0.0015$\pm$0.0009 & 0.0000$\pm$0.0001 &0.0003$\pm$0.0003 \\
$e\sin{\omega}$  &-0.0018$\pm$0.0019& -0.0014$\pm$0.0019 &-0.0021$\pm$0.0021 \\
${\rm reflected~light}_A$, mag  &0.010$\pm$0.005 & -0.003$\pm$0.001 &0.005$\pm$0.002 \\
${\rm reflected~light}_B$, mag  &0.010$\pm$0.005 & 0.003$\pm$0.001 &0.010$\pm$0.002 \\
$L_3$, per cent & $4.33\pm9.87$& $12.04\pm1.95$ & $29.49\pm3.93$\\
$S_0$, mag & $13.586\pm0.006$& $-0.009\pm0.002$& $-5.546\pm0.002$\\
$P$,d  &1.217783$\pm$0.000001 & 1.217785$\pm$0.000002 &1.217785$\pm$0.000002 \\
$t_0$, HJD d& 2458808.805418$\pm$0.000334 & 2458808.805166$\pm$0.000972 & 2458808.805001$\pm$0.000977 \\
$K_A,\,\kms$ &122.95$\pm$0.21 & 122.99$\pm$0.20&122.99$\pm$0.25 \\
$K_B,\,\kms$ &142.39$\pm$0.33 & 142.44$\pm$0.35&142.44$\pm$0.34 \\
$\gamma_A,\,\kms$ &-17.15$\pm$0.13 &-17.04$\pm$0.11 &-17.06$\pm$0.13 \\
$\gamma_B,\,\kms$ &-17.46$\pm$0.24 & -17.59$\pm$0.22 &-17.57$\pm$0.22 \\
\hline
derived\\
$L_B/L_A$  &0.333$\pm$0.075 & 0.576$\pm$0.107 &0.540$\pm$0.223 \\
$e$  &0.002$\pm$0.001 & 0.001$\pm$0.001 &0.002$\pm$0.001 \\
$\omega^{\circ}$ &309.67$\pm$123.88 & 266.45$\pm$86.00 &278.73$\pm$90.11 \\
$a,\,R_\odot$   &6.425$\pm$0.022 & 6.457$\pm$0.010 &6.466$\pm$0.016 \\
$q$  &0.863$\pm$0.003 & 0.863$\pm$0.003 &0.863$\pm$0.003 \\
$M_A,\,M_\odot$   &1.288$\pm$0.014 & 1.307$\pm$0.007 &1.313$\pm$0.010 \\
$M_B,\,M_\odot$   &1.112$\pm$0.011 & 1.129$\pm$0.005 &1.133$\pm$0.008 \\
$R_A,\,R_\odot$   &1.516$\pm$0.043 & 1.405$\pm$0.052 &1.424$\pm$0.102\\
$R_B,\,R_\odot$   &1.012$\pm$0.077 & 1.219$\pm$0.060 &1.196$\pm$0.123 \\
$\logg_A$, cgs  &4.186$\pm$0.025 & 4.259$\pm$0.033 &4.249$\pm$0.065 \\
$\logg_B$, cgs  &4.474$\pm$0.057 & 4.319$\pm$0.042 &4.337$\pm$0.086 \\
\hline
    \end{tabular}
    \label{tab:lcorbit}
\end{table*}

\begin{table}
    \centering
    \caption{Spectral parameters from multiple-epoch fitting.}
    \begin{tabular}{lcc}
\hline
Parameter & Star A & Star B\\
\hline
\multicolumn{3}{l}{short epoch's spectra with free $\feh$ }\\
$\gamma,\, \kms$ & \multicolumn{2}{c}{$-17.61 \pm 0.28$}\\
$q_{\rm dyn}$ & \multicolumn{2}{c}{$0.867\pm0.005$}\\
$\teff$, K & 6104$\pm$13 & 5980$\pm$25 \\
$\logg$, cgs & 3.86$\pm$0.02 & 4.07$\pm$0.02\\
$\feh$, dex & $-0.29\pm0.01$  &  0.02$\pm$0.02\\
$\vsini,\,\kms$ & 58$\pm$1  & 50$\pm$1\\
\multicolumn{3}{l}{stacked spectra with free $\feh$ }\\
$\gamma,\, \kms$ & \multicolumn{2}{c}{$-17.34 \pm 0.09$}\\
$q_{\rm dyn}$ & \multicolumn{2}{c}{$0.866\pm0.003$}\\
$\teff$, K & 6107$\pm$4 & 5941$\pm$9 \\
$\logg$, cgs & 3.86$\pm$0.01 & 4.07$\pm$0.01 \\
$\feh$, dex & $-0.36\pm0.01$  &  -0.05$\pm$0.01 \\
$\vsini,\,\kms$ & 59$\pm$1  & 51$\pm$1\\
\multicolumn{3}{l}{short epoch's spectra, assuming same $\feh$}\\
$\gamma,\, \kms$ & \multicolumn{2}{c}{$-17.54 \pm 0.28$}\\
$q_{\rm dyn}$ & \multicolumn{2}{c}{$0.865\pm0.005$}\\
$\teff$, K & 6177$\pm$9 & 5820$\pm$15 \\
$\logg$, cgs & 3.88$\pm$0.02 & 4.00$\pm$0.02\\
$\feh$, dex & \multicolumn{2}{c}{$-0.19\pm0.01$}\\
$\vsini,\,\kms$ & 59$\pm$1  & 50$\pm$1\\
\multicolumn{3}{l}{stacked spectra, assuming same $\feh$}\\
$\gamma,\, \kms$ & \multicolumn{2}{c}{$-17.32 \pm 0.09$}\\
$q_{\rm dyn}$ & \multicolumn{2}{c}{$0.866\pm0.002$}\\
$\teff$, K & 6140$\pm$3 & 5827$\pm$6 \\
$\logg$, cgs & 3.89$\pm$0.01 & 4.02$\pm$0.01\\
$\feh$, dex & \multicolumn{2}{c}{$-0.27\pm0.01$}\\
$\vsini,\,\kms$ & 59$\pm$1  & 51$\pm$1\\

\hline
    \end{tabular}
    \label{tab:final}
\end{table}

\subsection{SED fitting}
\begin{figure}
	\includegraphics[width=\columnwidth]{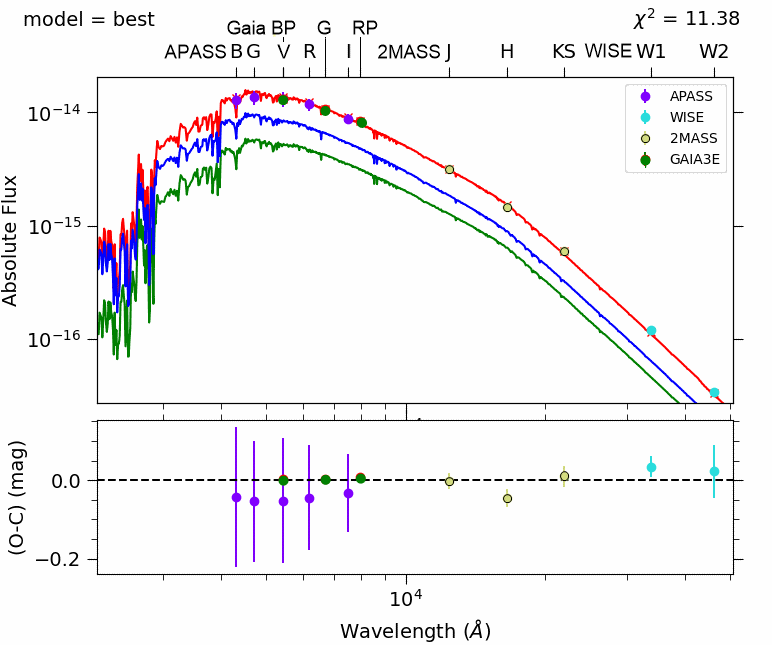}
    \caption{SED fitting with {\sc speedyfit}. The top panels shows observations and spectral energy distribution of the system (red line), primary (blue line) and secondary (green line).  The fit residuals are shown on the bottom panel. }
    \label{fig:sed}
\end{figure}

 Spectral energy distribution (SED) can provide independent estimate of system's parameters. We use published photometry and use out-of-eclipse measurements, when it is possible. 
To fit the SED we apply the {\sc speedyfit} package\footnote{https://github.com/vosjo/speedyfit}, which uses \cite{kurucz} models with
a Markov chain Monte-Carlo approach to find the global minimum and determine the errors on the fit parameters. We set constraints on the solution using the Gaia DR3 parallax $\varpi=0.8915 \pm0.0177$ mas, masses and radii from LC solution. Errors on the constraints are included as priors in the Bayesian fitting process and are thus propagated to the final results. A more detailed description of the fitting process is given in \citet{Vos2017}. {\sc speedyfit} doesn't support ZTF and TESS filters, therefore we don't use them.  We show resulting fit in Figure~\ref{fig:sed} and list parameters in Table~\ref{tab:sed}. $\teff$ values are slightly differ from spectroscopic ones, with residuals are within errors.  

\begin{table}
    \centering
        \caption{SED fitting results and observed photometry from Gaia DR3 \protect\cite{gaia3}, APASS \protect\cite{apass}, 2MASS \protect\cite{2mass} and WISE \protect\cite{wise}. Gaia DR3 and WISE values were computed as mean and standard deviation of out-of-eclipse parts of the time-series.}
    \begin{tabular}{lc}
        \hline
        Observations (mag):\\
        GDR3 $G$ & $13.416\pm0.008$\\
        GDR3 $BP$ & $13.715\pm0.009$\\
        GDR3 $RP$ & $12.949\pm0.007$\\
        APASS $B$ & $14.22\pm0.18$\\
        APASS $V$ & $13.61\pm0.16$\\
        APASS $G$ & $13.90\pm0.15$\\
        APASS $R$ & $13.45\pm0.13$\\
        APASS $I$ & $13.32\pm0.10$\\
        2MASS $J$ & $12.46\pm0.02$\\
        2MASS $H$ & $12.22\pm0.02$\\
        2MASS $K_s$ & $12.13\pm0.03$\\
        WISE $W1$ & $12.08\pm0.03$\\
        WISE $W2$ & $12.11\pm0.07$\\
        \hline
        Fitted parameters:\\
        $d$, pc& {$1107_{-11}^{+12}$}\\
        $E(B-V)$, mag& {$0.005_{-0.029}^{+0.039}$}\\
        ${\teff}_{A,B}$, K& $6006_{-172}^{+294},\,5832_{-199}^{+316}$\\
        \hline
        
    \end{tabular}

    \label{tab:sed}
\end{table}

\subsection{Selection of SB2 candidates.}
\begin{figure}
    \centering
    \includegraphics[width=\columnwidth]{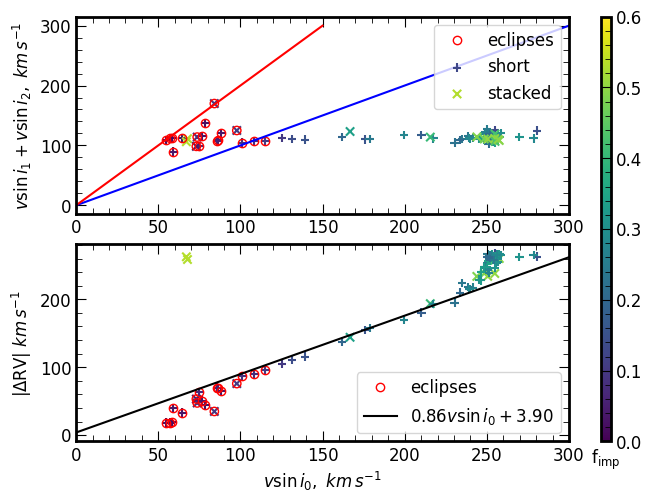}
    \caption{$\vsini_0$ values from the single-star fit versus $\vsini_1+\vsini_2$ from the binary fit (top panel) and versus $|\Delta\rv|$ (bottom panel). Blue and red solid lines show the functions $\vsini_1+\vsini_2=\vsini_0$ and $\vsini_1+\vsini_2= 2\vsini_0$ respectively. Black line shows best linear fit for a binary sample from \protect\cite{cat22}. The data for spectra taken during eclipses are highlighted by the red open circles.}
    \label{fig:sel}
\end{figure}
 
 In Figure~\ref{fig:sel} we explore how the usage of the short epoch spectra can affect new SB2 selection method from \cite{cat22}. This method is based on $\vsini$ estimate from the single-star fit, and can detect SB2 with smaller $|\Delta\rv|$ than traditional methods, based on cross-correlation functions. We can see that all short epoch spectra below the blue line can be selected. All of them were observed in out of the eclipse phases. Derived sum of rotational velocities is nearly constant for them. Unlike stacked spectra we don't have any short epoch spectra, where single-star model fitted only one component. Two such stacked spectra are clearly seen in the bottom panel near upper left corner. Also we can clearly see, that linear relation derived in \cite{cat22} for $|\Delta\rv|$ vs  $\vsini_0$ is correct for short epoch spectra, observed in out of the eclipse phases, except for region with $\vsini_0>240~\kms$, where $\vsini_0$ ``saturates".   

\section{Discussion}
\label{discus}

\begin{table}
    \centering
    \caption{Literature data compilation from ZTF \protect\cite{ztf_var}, ASAS-SN \protect\cite{asassnv}, Gaia DR3
        FLAME and Multiple Star Classifier (MSC) \protect\cite{gaia3}, UNIDAM \protect\cite{unidam} and \protect\cite{xiong}. Gaia DR3 values provided with 16 and 84 percentiles. (*) assuming single star.}
    \begin{tabular}{lcl}
        \hline
        Parameter & Value & reference\\
        \hline
        $P$, d& {$1.2177044$}& ZTF\\
        $P$, d& {$1.217726$}& ASAS-SN\\
        $d$, pc& {$1080_{10}^{1320}$}& GDR3 MSC\\
        $A_G$, mag& {$0.3131_{0.}^{0.9389}$}& GDR3 MSC\\
        \hbox{[Fe/H]}, dex& {$0.13_{-0.10}^{0.5}$}&GDR3 MSC\\
        ${\teff}_{A,B}$, K& $6127_{5877}^{7821}$, $5885_{4018}^{6558}$&GDR3 MSC\\
        $\logg_{A,B}$, cgs& $4.28_{4.08}^{4.52}$, $4.57_{4.13}^{5.2}$&GDR3 MSC\\
        ${\teff}_{A,B}$, K& $6193\pm39,\,5654\pm57$& Xiong\\
        $\logg_{A,B}$, cgs& $4.09\pm0.08,\,4.64$& Xiong\\
        $\feh_{A,B}$, dex& $-0.250\pm0.110$& Xiong\\
        $M_{A,B},\,M_\odot$ & $1.407\pm0.011,\,1.214\pm0.011$ & Xiong\\
        $R_{A,B},\,R_\odot$ & $1.501\pm0.050,\,0.868\pm0.027$ & Xiong\\
        $\vsini_{A},\,\kms$ & $91.80$  & Xiong\\
        $q$ & $0.862\pm0.015$  & Xiong\\
        age ${\rm t}$, Gyr& {$5.664_{4.983}^{6.317}$}& GDR3 FLAME*\\
        age $\log{\rm t}$, yr& {$9.572 \pm 0.246$}& UNIDAM*\\
        \hline
        
    \end{tabular}

    \label{tab:lit}
\end{table}

\subsection{System parameters}

The physical parameters of the components suggests that the system consists two fast rotating stars, which are heavier and larger than the Sun. They form a close system with a circular orbit, slightly inclined to the line of sight, therefore we can see partial eclipses. Both components are on the main sequence and mass transfer is not started yet, as both components are smaller than their Roche lobes. The spectroscopic solution is computed with an assumption of the same $\feh$ for both components have $R_B/R_A=0.81$ which is very close to the value $R_B/R_A=0.868\pm0.076$ from the TESS 44 LC fit. The relatively large projected rotational velocities measured from the spectra are in perfect agreement with ${\vsini}_{{\rm sync}\,A,B}=57.8\pm2.0,\,50.0\pm2.5\,\kms$ computed using TESS LC fit parameters, suggesting spin-orbit synchronisation.
\par
Surface gravities derived by \texttt{JKTEBOP} are significantly larger than the spectroscopic ones. There is a bias of $\Delta\logg= 0.3-0.4$ dex, which is larger than the expected $\logg$ uncertainties given in Section~\ref{sec:err}. Similar bias was observed in \cite{Kovalev19} for open cluster NGC~2243, where stars near the turn-off have spectroscopic $\logg$ smaller than $\logg$ from the PARSEC isochrone \citep{marigo2017}. Thus such a bias can be related to the inaccuracy of the spectral models.  Recent study by \cite{chen2022} also noted underestimation of the $\logg$ in LAMOST-MRS data when they are fitted by the binary model. Note that difference of spectroscopic surface gravities $\logg_B-\logg_A=0.12$ is comparable to difference ($\logg_B-\logg_A=0.06$) from the TESS LC fit, because we use it to scale the relative contribution of the components in the spectrum. 
\par
We explore the previously reported measurements for this system and collect them in Table~\ref{tab:lit}. Orbital periods from ZTF and ASAS-SN~V are in good agreement with our estimations, although they are slightly shorter. The recent Gaia DR3 presents very interesting estimates by Multiple Star Classifier (MSC) \cite{gaia3}, where they tried to model observed low resolution $BP,\,RP$ spectra using two stellar components. These estimates agree very well with our $\teff$ from the multiple-spectra fit and SED analysis and $\logg$ from the TESS 44 dataset, although our $\feh$ is 0.3-0.4 dex smaller. This is a well-behaved solution based on goodness-of-fit score \texttt{logposterior\_msc}=495\footnote{\url{https://gea.esac.esa.int/archive/documentation/GDR3/Data_analysis/chap_cu8par/sec_cu8par_apsis/ssec_cu8par_apsis_msc.html}}. We also retrieve all their 100 Monte Carlo samples and plot them in Figure~\ref{fig:msc}.
\par
\cite{xiong} also used LAMOST-MRS, ZTF, ASAS-SN data to analyse this system with \texttt{PHOEBE}. For primary star in the system they got $M_A$ and $R_A$ larger than our estimates, but their ${\teff}_A$ agrees with our value. For the secondary star their $M_B$ is higher, but $R_B$ and ${\teff}_B$ are significantly smaller than our estimates. Their $\feh$ for the system agrees with our estimate from the stacked spectra. We think that the disagreement is due to their usage of less precise LC datasets and the usage of spectral parameters from the second minimum observations for initialisation. They likely underestimated the orbital inclination, which led to much larger masses.
\par
 We take parameters from TESS~44 fit by \texttt{JKTEBOP} as a final solution as this LC has much better quality and greater quantity than other two. For $\teff$ and $\feh$ we choose the result from multiple-epochs fit of short epochs with the same $\feh$ for both components. Additionally we use {\sc PHOEBE} to verify our results and find that ZTF $r$ LC can be consistently fitted together with other two TESS LCs, see Appendix~\ref{phoebefit}. 

\subsection{Age of the system}
\begin{figure}
    \centering
    \includegraphics[width=0.45\textwidth]{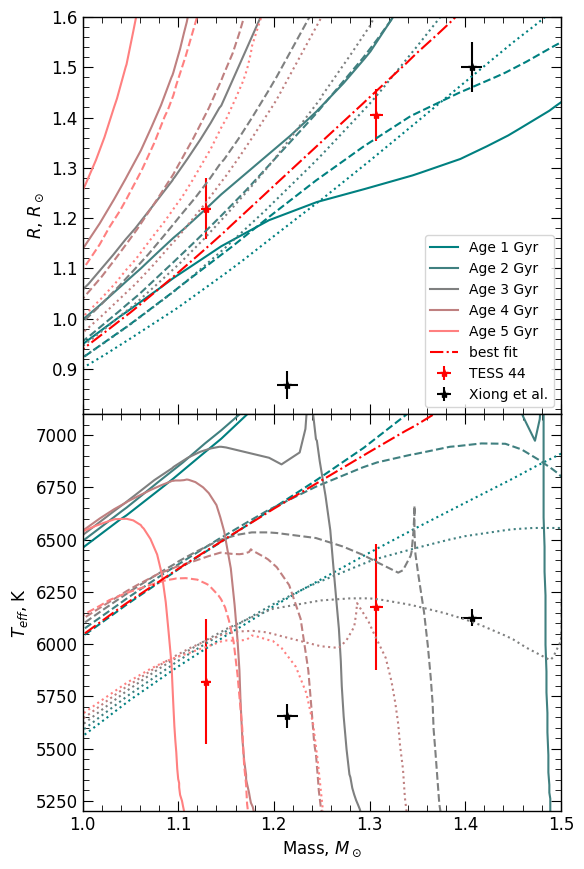}
    \caption{The grid of PARSEC isochrones computed for $\feh=-0.6$ (solid lines), $-0.2$ (dashed lines), $0.2$ dex (dotted lines) and five ages in range from 1 to 5 Gyr. Our best fit ($\feh=-0.19$ dex, age 1.57 Gyr) isochrone is also shown (red dot-dashed line), together with our best estimates and parameters from \protect\cite{xiong}.}
    \label{fig:my_age}
\end{figure}

We use the grid of  {\sc PARSEC} isochrones \citep{marigo2017}, computed at metallicity $\feh=-0.19$ dex, together with masses and radii from all RP simulations for the TESS 44 dataset to derive the system's age, assuming that the two stars evolve separately, but have the same age. All fitted age solutions have median $\log{\rm t}=9.195_{9.175}^{9.230}$ yr (1.56 Gyr) with 16 and 84 percentiles, which is significantly younger than the Sun. This is very different from the single star age values ${\rm t}=5.664_{4.983}^{6.317}$ Gyr from GDR3 FLAME \citep{gaia3} and $\log{\rm t}=9.572 \pm 0.246$ (3.73 Gyr) from \citep{unidam}.
The theoretical circularisation time for an orbit with a given $q$ and $P$, using Formula 6.2 from \cite{zahn71}: $\log{{\rm t}_{\rm circ}}=9.47$ yr (2.95 Gyr) is larger than our estimate of the system's age, therefore tidal friction was very efficient and possibly was also accompanied by strong magnetic interactions. 
However we should note, that theoretical models have difficulty describing this system. In Figure~\ref{fig:my_age} we plot a grid of {\sc PARSEC} isochrones computed for $\feh=-0.6,\,-0.2,\,0.2$ dex and five ages in a range from 1 to 5 Gyr. Our best fit isochrone is also shown, together with our best estimates and parameters from \cite{xiong}. Generally, theoretical models are steeper in the mass-radius plane than our estimates and therefore the best-fit isochrone goes through the upper errorbar for the primary and lower errorbar for the secondary. In the lower panel with the mass-$\teff$ plane, you can see that theoretical values of $\teff$ are significantly larger than our $\teff$ estimates at corresponding masses. Only isochrones with significant super-solar metallicity will be able to match them. Therefore our results suggest that PARSEC isochrones are unable to model this close dEB system properly and our estimate of age is not reliable.     
\par
 A similar discrepancy was found by \cite{zzuma} for ZZ UMa, where {\sc PARSEC} isochrones were not able to match all parameters in the system. For the secondary component the measured radius ($R=1.0752 \pm 0.0046~R_\odot$) is too large and the temperature ($\teff=5270\pm90$ K) is too low in comparison with isochrone predictions, that well-match the primary component. \cite{zzuma} suggest that this is case of \textit{radius discrepancy}, which is well-known phenomena for the late-type dwarfs.
\par
Recently \citet{2018MNRAS.479.1953D} employed a set of well studied dEBs to test {\sc PARSEC} isochrones, using a Bayesian framework to infer masses, distances and ages, taking radii, $\teff$ and $\feh$ as input.  They found that {\sc PARSEC} models systematically underestimate masses, although discrepancies  are  not  so  important for the main sequence. 
For example, the close dEB system UZ Dra with similar parameters to J064726.39+223431.6 ($P=3.261$ d, $R_{A,B}=1.31\pm0.03,\,1.15\pm0.02~R_\odot$, ${\teff}_{A,B}=6209\pm114,\,5984 \pm110$ K, $M_{A,B}=1.34\pm0.02,\,1.23\pm0.02~M_\odot$, no estimate of $\feh$ yet) they inferred $M^{\rm fit}_{A,B}=1.13\pm0.09,1.04\pm0.09~M_\odot$, assuming solar metallicity for this system. 
\par
Additionally we check another two sets of theoretical models by \cite{dotter16} and \citet{2019A&A...628A..29C}, but find no improvement relative to {\sc PARSEC}. 

\section{Conclusions}
\label{concl}
We present a study of the dEB SB2 J064726.39+223431.6 system using spectra from the LAMOST-MRS survey and photometrical data. We use full-spectrum fitting to derive radial velocities and spectral parameters. The orbital solution and light curves from ZTF and TESS suggest that it is a close pair on a circular orbit. We have measured the masses of the stars to accuracies of 1 per cent and the radii to accuracies of 5 per cent. The system shows strong evidence for spin-orbit synchronisation. We find a discrepancy between spectroscopic $\logg$ and $\logg$ calculated using parameters from the LC fitting. We found that $\feh$ derived from the stacked spectra is $0.07$ dex smaller than $\feh$ calculated using short epochs. Theoretical isochrones also unable to model the measured mass-$\teff$ relations at derived metallicity. The derived age of the system indicates that both components are still on the main sequence, although it is much shorter than orbit's circularisation timescale. 

\section*{Acknowledgements}
 We are grateful to the anonymous referee for a constructive report.
We thank Hans B{\"a}hr for his careful proof-reading of the manuscript.
We thank Sa{\v s}a Iliji{\'c} for help with {\sc FD3} code.
MK is grateful to his parents, Yuri Kovalev and Yulia Kovaleva, for their full support in making this research possible. The work is supported by the Natural Science Foundation of China (Nos. 11733008, 12090040, 12090043, 11521303, 12125303, 12273057).
S. W. acknowledges support from the Youth Innovation Promotion Association of the CAS (id. 2019057).
Guoshoujing Telescope (the Large Sky Area Multi-Object Fiber Spectroscopic Telescope LAMOST) is a National Major Scientific Project built by the Chinese Academy of Sciences. Funding for the project has been provided by the National Development and Reform Commission. LAMOST is operated and managed by the National Astronomical Observatories, Chinese Academy of Sciences. The authors gratefully acknowledge the “PHOENIX Supercomputing Platform” jointly operated by the Binary Population Synthesis Group and the Stellar Astrophysics Group at Yunnan Observatories, Chinese Academy of Sciences. 
This research has made use of NASA’s Astrophysics Data System, the SIMBAD data base, and the VizieR catalogue access tool, operated at CDS, Strasbourg, France. It also made use of TOPCAT, an interactive graphical viewer and editor for tabular data \citep[][]{topcat}.  Funding for the
TESS mission is provided by NASA’s Science Mission directorate. This paper includes data collected by the TESS mission, which is publicly available from the Mikulski Archive for Space Telescopes (MAST). This work has made use of data from the European Space Agency (ESA) mission
{\it Gaia} (\url{https://www.cosmos.esa.int/gaia}), processed by the {\it Gaia}
Data Processing and Analysis Consortium (DPAC,
\url{https://www.cosmos.esa.int/web/gaia/dpac/consortium}). Funding for the DPAC
has been provided by national institutions, in particular the institutions
participating in the {\it Gaia} Multilateral Agreement.
\par
Based on observations obtained with the Samuel Oschin Telescope 48-inch and the 60-inch Telescope at the Palomar Observatory as part of the Zwicky Transient Facility project. ZTF is supported by the National Science Foundation under Grant No. AST-2034437 and a collaboration including Caltech, IPAC, the Weizmann Institute for Science, the Oskar Klein Center at Stockholm University, the University of Maryland, Deutsches Elektronen-Synchrotron and Humboldt University, the TANGO Consortium of Taiwan, the University of Wisconsin at Milwaukee, Trinity College Dublin, Lawrence Livermore National Laboratories, and IN2P3, France. Operations are conducted by COO, IPAC, and UW
This research has made use of the NASA/IPAC Infrared Science Archive, which is funded by the National Aeronautics and Space Administration and operated by the California Institute of Technology.

\section*{Data Availability}
The data underlying this article will be shared on reasonable request to the corresponding author.




\bibliographystyle{mnras}




\appendix

\section{Spectral models}
\label{sec:payne}
The synthetic spectra are generated using NLTE~MPIA online-interface \url{https://nlte.mpia.de} \citep[see Chapter~4 in][]{disser} on wavelength intervals 4870:5430 \AA~for the blue arm and 6200:6900 \AA ~for the red arm with spectral resolution $R=7500$. We use NLTE (non-local thermodynamic equilibrium) spectral synthesis for H, Mg~I, Si~I, Ca~I, Ti~I, Fe~I and Fe~II lines \citep[see Chapter~4 in][ for references]{disser}.  
\par
The grid of models (6200 in total) is computed for points randomly selected in a range of $\teff$ between 4600 and 8800 K, $\logg$ between 1.0 and 4.8  (cgs units), $\vsini$ from 1 to 300 $\kms$ and [Fe/H]\footnote{We used $\feh$ as a proxy of overall metallicity, abundances for all elements are scaled with Fe.} between $-$0.9 and $+$0.9 dex. The model is computed only if linear interpolation of the MAFAGS-OS\citep[][]{Grupp2004a,Grupp2004b} stellar atmosphere is possible for a given point in parameter space.  Microturbulence is fixed to $\Vmic=2~\kms$ for all models. The grid is randomly split on training (70\%) and cross-validation (30\%) sets of spectra, which are used to train \textit{The~Payne} spectral model \citep{ting2019}. The neural network (NN) consists of two layers of 300 neurons each with rectilinear unit (ReLU)\footnote{ReLU(x)=max(x,0)} activation functions. We train separate NNs for each spectral arm. The median approximation error is less than 1\% for both arms. We use output of \textit{The Payne} as single-star spectral model.   

\section{RV measurements}
 We provide RV measurements from the short epochs in Table~\ref{tab:rvs}.
\begin{table}
    \centering
    \caption{\label{tab:rvs} Radial velocity measurements. The asterisk (*) denotes datapoints which are not used in analysis with \texttt{GLS} and {\sc JKTEBOP}. We subtract 2400000.5 days from time values.}
    \begin{tabular}{lccc}
\hline
time &  ind. epoch & single  \\
HJD & RV$_{ B}$ , RV$_{A}$& RV \\
 d    & $\kms$ & $\kms$\\

\hline
 58806.744 &  98.54$\pm$2.07, -169.77$\pm$2.50 & -7.01  \\
 58806.760 & 103.66$\pm$1.71, -159.21$\pm$2.72 & -1.42  \\
 58806.776 &-159.98$\pm$1.02,  106.50$\pm$0.70 & -5.79  \\
 58806.792 &-158.66$\pm$1.12,  106.87$\pm$0.80 & 2.57  \\
 58806.808 &-158.83$\pm$0.97,  103.80$\pm$0.68 & -5.38  \\
 58806.825 &-155.77$\pm$1.06,  103.79$\pm$0.72 & -7.45  \\
 58806.841 &-152.48$\pm$1.11,  100.31$\pm$0.72 & -0.87  \\
 58806.857 &-151.16$\pm$1.10,   96.59$\pm$0.79 & -9.72  \\
 58820.774 & 124.71$\pm$0.89, -141.97$\pm$0.66 & -20.69  \\
 58820.783 & 124.27$\pm$0.97, -139.47$\pm$0.66 & -26.77  \\
 58820.792 & 123.54$\pm$0.92, -140.18$\pm$0.64 & -21.63  \\
 58820.802 & 121.96$\pm$1.00, -139.56$\pm$0.69 & -23.36  \\
 58820.811 & 122.94$\pm$1.00, -139.70$\pm$0.73 & -16.79  \\
 58820.820 & 123.32$\pm$0.92, -138.43$\pm$0.68 & -19.63  \\
 58820.830 & 121.06$\pm$1.01, -137.90$\pm$0.67 & -26.18  \\
 58835.736 & -40.86$\pm$2.70,    3.96$\pm$1.90*& -12.50  \\
 58835.753 & -45.46$\pm$1.58,   24.66$\pm$1.34*& -9.52  \\
 58835.769 & -64.15$\pm$1.63,   22.22$\pm$1.08*& -9.63  \\
 58835.785 &  30.49$\pm$0.91,  -74.37$\pm$1.36*& -10.97  \\
 58856.640 &-149.11$\pm$1.20,   96.22$\pm$0.84 & 0.62  \\
 58856.656 &-154.25$\pm$1.13,   98.60$\pm$0.87 & -8.63  \\
 58856.672 &-157.34$\pm$1.25,  104.05$\pm$0.90 & -2.07  \\
 58856.688 &-157.91$\pm$0.99,  105.83$\pm$0.71 & -12.86  \\
 58856.704 &-158.98$\pm$1.20,  105.49$\pm$0.84 & -6.12  \\
 58856.720 &-158.31$\pm$1.05,  106.23$\pm$0.83 & -12.57  \\
 58860.669 &  -9.49$\pm$2.36,  -27.99$\pm$3.65*& -16.65  \\
 58860.685 &   0.21$\pm$2.98,  -39.85$\pm$3.50*& -21.37  \\
 58860.701 &   9.52$\pm$3.42,  -41.39$\pm$2.49*& -20.35  \\
 58860.717 &  25.16$\pm$2.84,  -43.71$\pm$1.84*& -22.23  \\
 58883.558 &-154.79$\pm$0.85,  101.26$\pm$0.63 & -8.13  \\
 58883.574 &-151.01$\pm$0.91,   96.86$\pm$0.67 & -8.01  \\
 58883.590 &-145.40$\pm$1.23,   93.86$\pm$0.88 & -8.07  \\
 58883.606 &-140.24$\pm$1.00,   88.86$\pm$0.72 & -8.89  \\
 58883.623 &  83.10$\pm$0.72, -132.46$\pm$0.95 & -13.08  \\
 58889.536 &-155.40$\pm$0.78,  100.59$\pm$0.59 & -6.28  \\
 58889.552 &-157.68$\pm$0.91,  103.44$\pm$0.63 & -9.97  \\
 58889.568 &-158.45$\pm$1.17,  106.64$\pm$0.86 & -3.32  \\
 58889.585 &-157.72$\pm$1.20,  105.39$\pm$0.87 & -5.89  \\
 58889.601 &-159.82$\pm$1.15,  106.04$\pm$0.86 & -10.49  \\
 58889.617 &-157.74$\pm$1.38,  104.89$\pm$1.03 & -7.40  \\
 58889.634 &-157.47$\pm$1.70,  103.59$\pm$1.28 & -7.53  \\
 58919.483 & 120.69$\pm$1.05, -134.31$\pm$0.73 & -25.05  \\
 58919.500 & 115.05$\pm$0.93, -131.58$\pm$0.65 & -23.05  \\
 58919.516 & 115.08$\pm$1.11, -127.50$\pm$0.75 & -13.90  \\
 58919.533 & 103.32$\pm$1.31, -120.86$\pm$0.84 & -25.18  \\
 58919.548 & 100.36$\pm$1.10, -116.91$\pm$0.76 & -17.02  \\
 \hline
\end{tabular}
\end{table}

\begin{table}\centering\contcaption{}
\begin{tabular}{lcc}
\hline
 &  ind. epoch & single  \\
HJD & RV$_{ B}$ , RV$_{A}$&RV \\
 d   &  $\kms$ & $\kms$\\
\hline
 59149.811 &  42.48$\pm$1.03,  -67.68$\pm$0.75*& -26.09  \\
 59149.827 &  29.63$\pm$1.34,  -61.11$\pm$0.89*& -27.31  \\
 59149.843 & -63.52$\pm$1.05,    1.57$\pm$1.21*& -25.46  \\
 59149.859 &  24.68$\pm$2.57,  -39.14$\pm$1.14*& -24.76  \\
 59149.876 &  -3.04$\pm$2.01,  -36.29$\pm$1.66*& -21.09  \\
 59149.892 & -28.63$\pm$2.69,  -10.16$\pm$1.36*& -15.98  \\
 59149.905 & -25.12$\pm$3.82,   -5.62$\pm$2.65*& -13.35  \\
 59180.759 &-134.31$\pm$1.42,   84.06$\pm$1.00 & -7.81  \\
 59180.775 &  78.87$\pm$1.14, -130.97$\pm$1.68 & -15.45  \\
 59180.791 &-121.85$\pm$1.40,   73.23$\pm$1.03 & -14.22  \\
 59180.807 &-112.51$\pm$1.68,   67.83$\pm$1.22 & -4.65  \\
 59180.824 & -97.08$\pm$1.62,   57.49$\pm$1.27 & -14.16  \\
 59189.734 & 120.61$\pm$1.38, -136.89$\pm$0.94 & -22.00  \\
 59189.750 & 122.79$\pm$1.66, -139.65$\pm$1.24 & -16.01  \\
 59189.767 & 125.09$\pm$1.54, -141.53$\pm$1.06 & -13.64  \\
 59189.783 & 125.75$\pm$1.80, -136.33$\pm$1.12 & -12.49  \\
 59235.592 &-121.19$\pm$0.94,   71.44$\pm$0.65 & -9.07  \\
 59235.609 &-109.18$\pm$1.12,   60.83$\pm$0.72 & -12.97  \\
 59235.625 &  56.43$\pm$0.75, -101.27$\pm$1.09 & -13.91  \\
 59235.641 &  47.93$\pm$0.71,  -89.90$\pm$1.06 & -14.00  \\
 59235.664 & -79.47$\pm$1.01,   36.19$\pm$0.75*& -14.24  \\
 59235.680 &  27.45$\pm$0.92,  -68.23$\pm$1.29*& -9.16  \\
 59264.548 &-133.20$\pm$0.93,   82.00$\pm$0.67 & -10.98  \\
 59264.564 &-140.56$\pm$0.82,   87.29$\pm$0.57 & -15.60  \\
 59264.580 &-146.59$\pm$0.76,   93.68$\pm$0.54 & -10.64  \\
 59264.597 &-150.17$\pm$0.78,   97.62$\pm$0.59 & -8.63  \\
\hline
\end{tabular}
    
\end{table}

\section{Verification with PHOEBE}
\label{phoebefit}
We use \texttt{PHOEBE}\citep{phoebe} to verify our results, since it allows one to fit multiple LC datasets simultaneously, and updates limb darkening coefficients during the fitting. For all three LCs we convert magnitudes to fluxes and divide them by the median value. We also subtract the best fit $t_0$ from the timescale in all datasets. We use the default Nelder-Mead optimiser and default ``ck2004" atmospheres. Eccentricity, period and mass ratio were fixed, thus we fit for $R_{\rm A,B},~{\teff}_{\rm A,B},~\gamma,~a\sin{i},~i$ and the third light contribution for three LC datasets. 

\begin{figure}
	\includegraphics[width=\columnwidth]{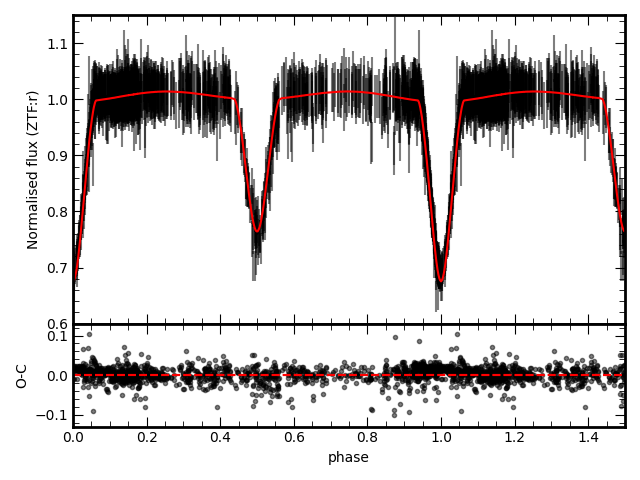}
	\includegraphics[width=\columnwidth]{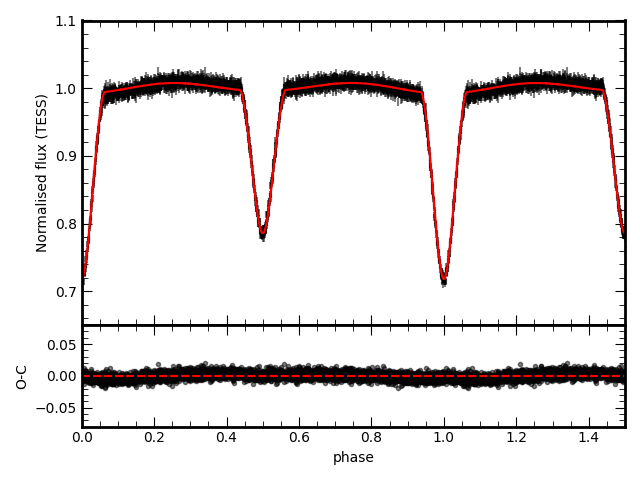}
	\includegraphics[width=\columnwidth]{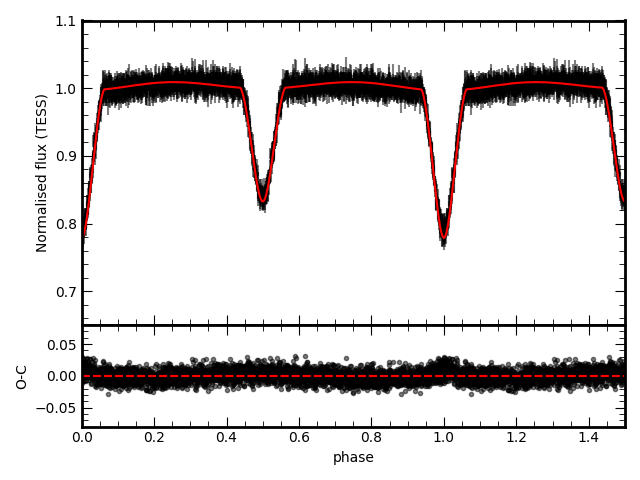}
	\caption{The best fit {\sc PHOEBE} model for ZTF $r$ (top), TESS 44 (middle) and TESS 45 (bottom) LCs. }
    \label{fig:phoebe}
\end{figure}

We present fitting results in Figure~\ref{fig:phoebe} and Table~\ref{tab:phoebe}. The best fit {\sc PHOEBE} model fits all three LCs well since the residuals (O-C) are insignificant.  All parameters agree with {\sc JKTEBOP} values for TESS datasets, therefore our usage of fixed limb darkening coefficients is reasonable. The best fit ${\teff}_B$ of the secondary agrees with the spectroscopic estimate, but ${\teff}_A$ is $250$ K higher. The computed $\vsini$ are in excellent agreement with spectroscopic values.
We don't use any sampling techniques to estimate errors as \texttt{PHOEBE} calculations are very computationally expensive in comparison with \texttt{JKTEBOP}, but we think uncertainties will be slightly smaller than ones, provided by {\sc PHOEBE} in \cite{xiong}.

\begin{table}
    \centering
    \caption{ Parameters derived by \texttt{PHOEBE} when fitting all three LCs simultaneously}.
    \begin{tabular}{lcc}
\hline
Parameter & Star A & Star B\\
\hline
fixed\\
$q$ & \multicolumn{2}{c}{$0.87$}\\
$P$, d& \multicolumn{2}{c}{$1.21778$}\\
$e$ & \multicolumn{2}{c}{$0.00$}\\
\hline
\multicolumn{3}{l}{fitted }\\
$\gamma,\, \kms$ & \multicolumn{2}{c}{$-17.25$}\\
$i\degr$ & \multicolumn{2}{c}{$81.63$}\\
$a\sin{i},\,R_\odot$ & \multicolumn{2}{c}{$6.421$}\\
$R_{\rm equiv},\,R_\odot$ & $1.427$ & $1.219$\\
$\teff$, K & $6414$ & $5873$ \\
$L_3$ ZTF $r$, per cent & \multicolumn{2}{c}{$0.0$}\\
$L_3$ TESS 44, per cent & \multicolumn{2}{c}{$12.0$}\\
$L_3$ TESS 45, per cent & \multicolumn{2}{c}{$30.4$}\\
\hline
derived\\
$M,\,M_\odot$ & 1.323 & 1.151\\
$\logg$, cgs & 4.251 & 4.327\\
$\vsini,\,\kms$ & 58.67  & 50.10\\
\hline
    \end{tabular}
    \label{tab:phoebe}
\end{table}

\begin{figure}
	\includegraphics[width=\columnwidth]{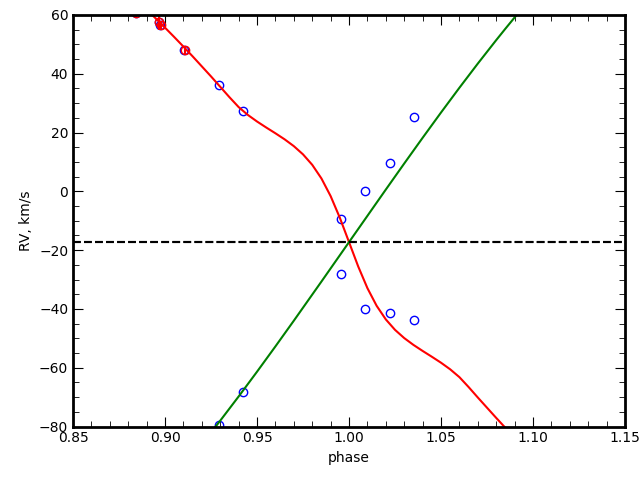}
	\includegraphics[width=\columnwidth]{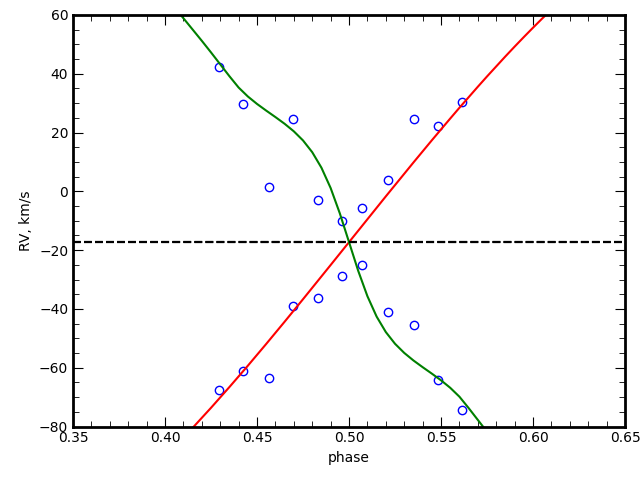}
	\caption{Radial velocities measured during eclipses together with best fit {\sc PHOEBE} model for primary (red line) and secondary (green line) components. Blue datapoints were not used in the fit.}
    \label{fig:rm}
\end{figure}

By phase-folding the whole RV dataset we find that several spectra were taken exactly during eclipses.
In Figure \ref{fig:rm} we show RVs measured from these spectra together with the flux-weighted RV curves computed by {\sc PHOEBE} using the best parameters. We can see that RM effect is not very strong, but it can be barely seen in our RVs data for primary eclipse, therefore eclipse is not total. 

\section{Residual-permutation simulations}
\begin{figure}
    \includegraphics[width=\columnwidth]{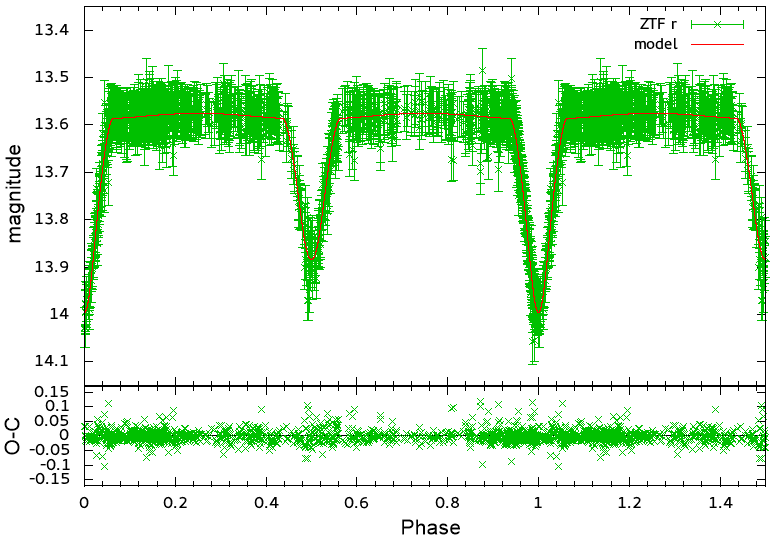}
    \caption{Phase-folded LC from ZTF $r$ fit with \texttt{JKTEBOP}. In the bottom panels we show fit residuals. }
    \label{fig:orbit_rg}
\end{figure}

\begin{figure}
	\includegraphics[width=\columnwidth]{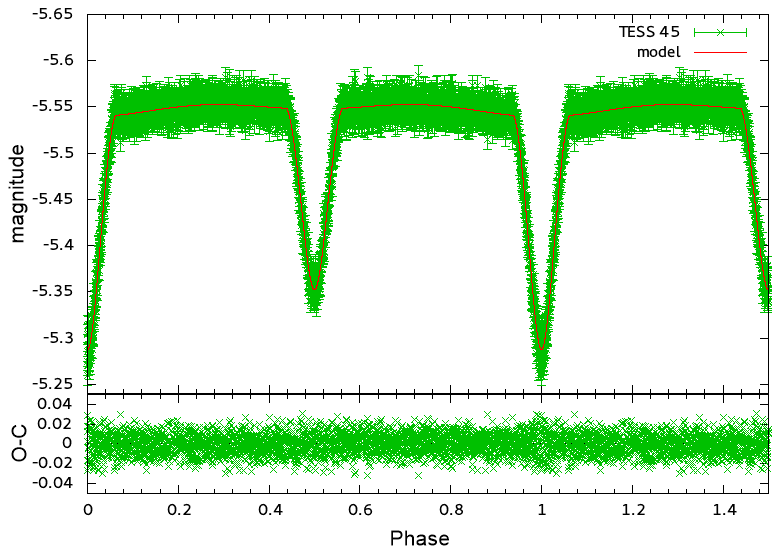}
    \caption{Phase-folded LC from TESS45 fit with \texttt{JKTEBOP}. In the bottom panels we show fit residuals. The magnitudes are not calibrated.}
    \label{fig:orbit_tess45}
\end{figure}

Fits for the ZTF~$r$ and TESS~45 LC datasets can be found in Figures \ref{fig:orbit_rg},\ref{fig:orbit_tess45}.

\begin{figure*}
    \includegraphics[width=\textwidth]{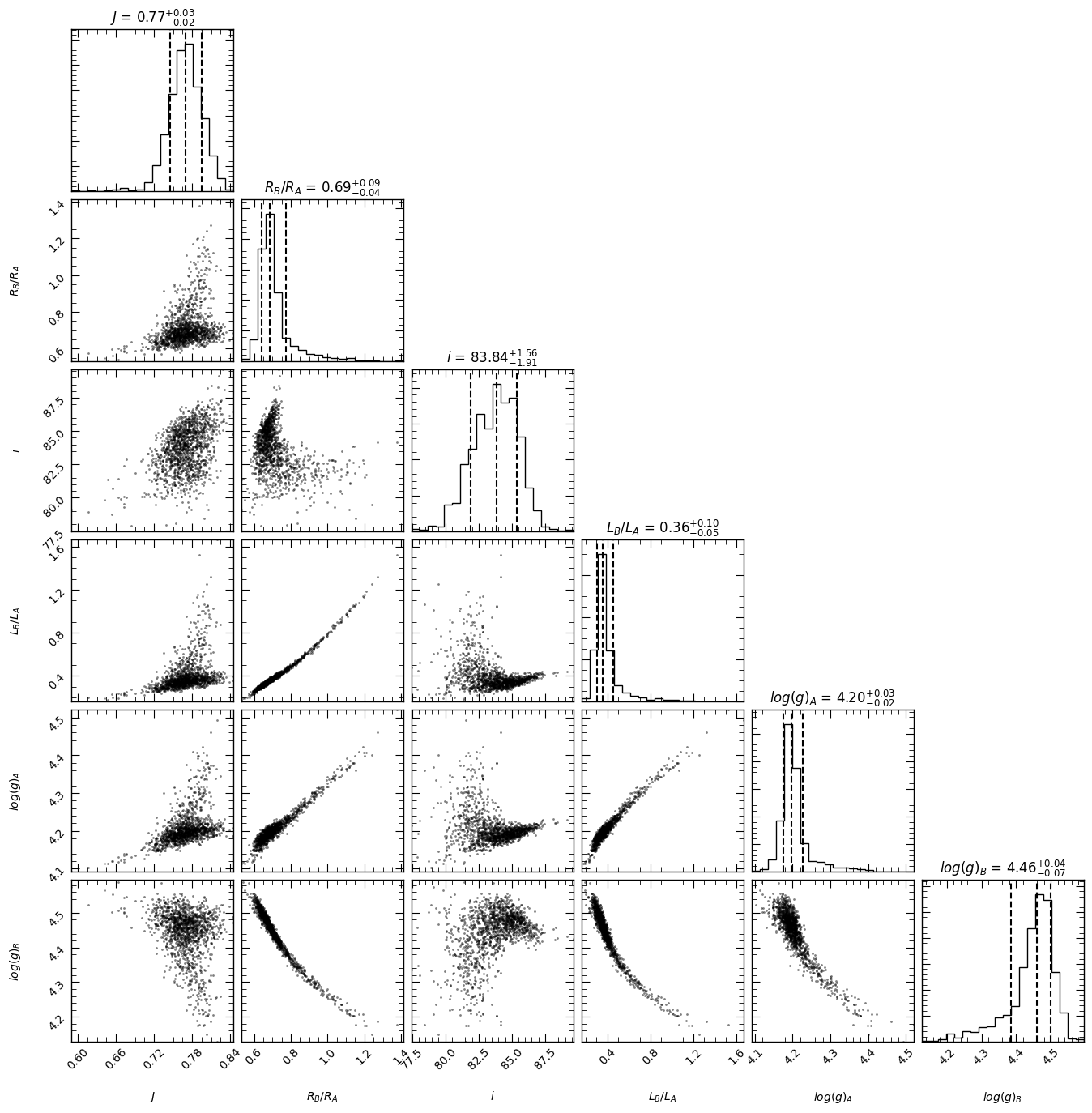}
    \caption{Corner plot for the residual-permutation simulations of ZTF $r$ solution. Titles show 16, 50 and 84 percentiles.}
    \label{fig:corner}
\end{figure*}

\begin{figure*}
    \includegraphics[width=\textwidth]{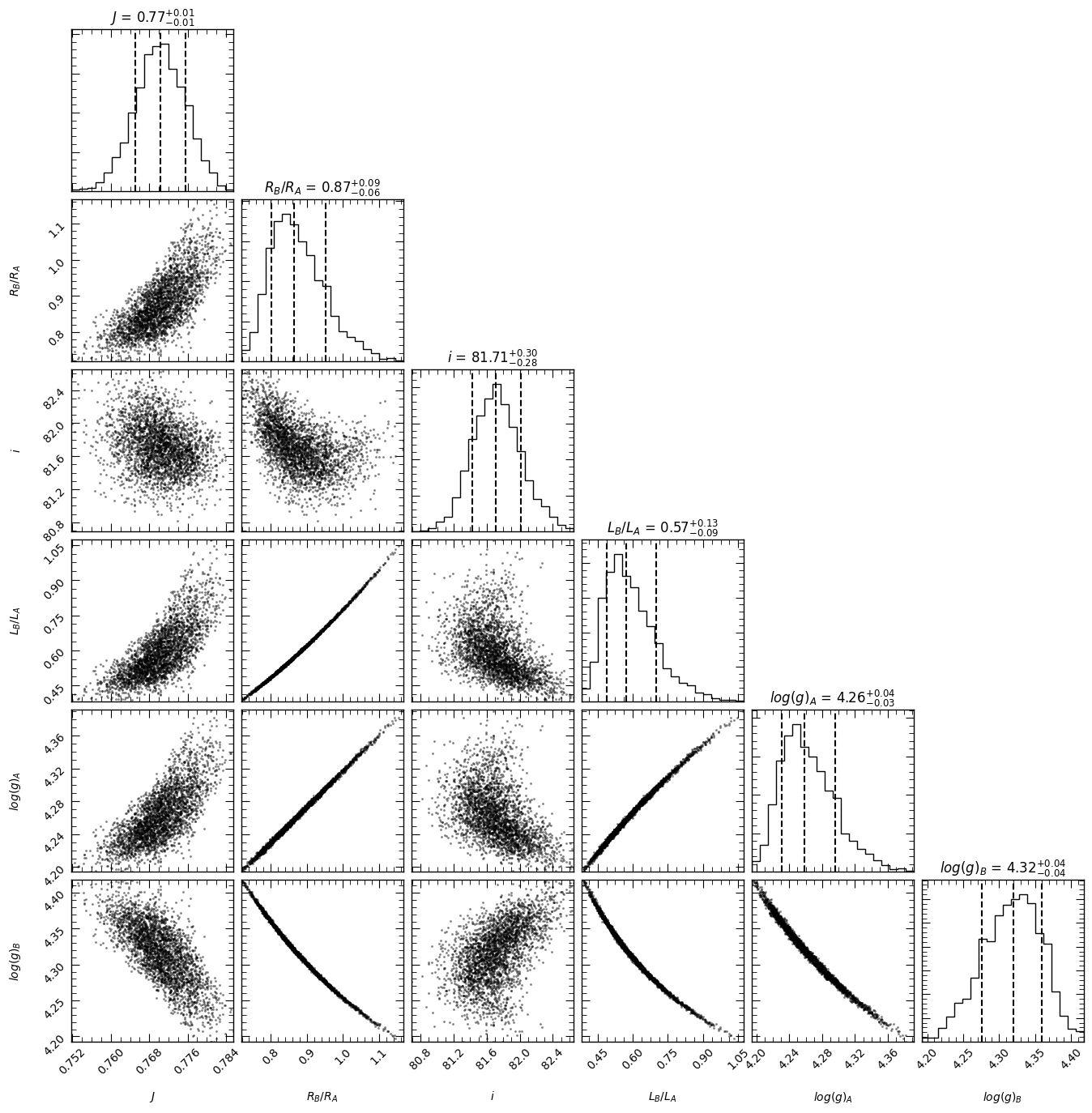}
    \caption{Corner plot for the residual-permutation simulations of TESS 44 solution. Titles show 16, 50 and 84 percentiles. }
    \label{fig:corner2}
\end{figure*}

\begin{figure*}
    \includegraphics[width=\textwidth]{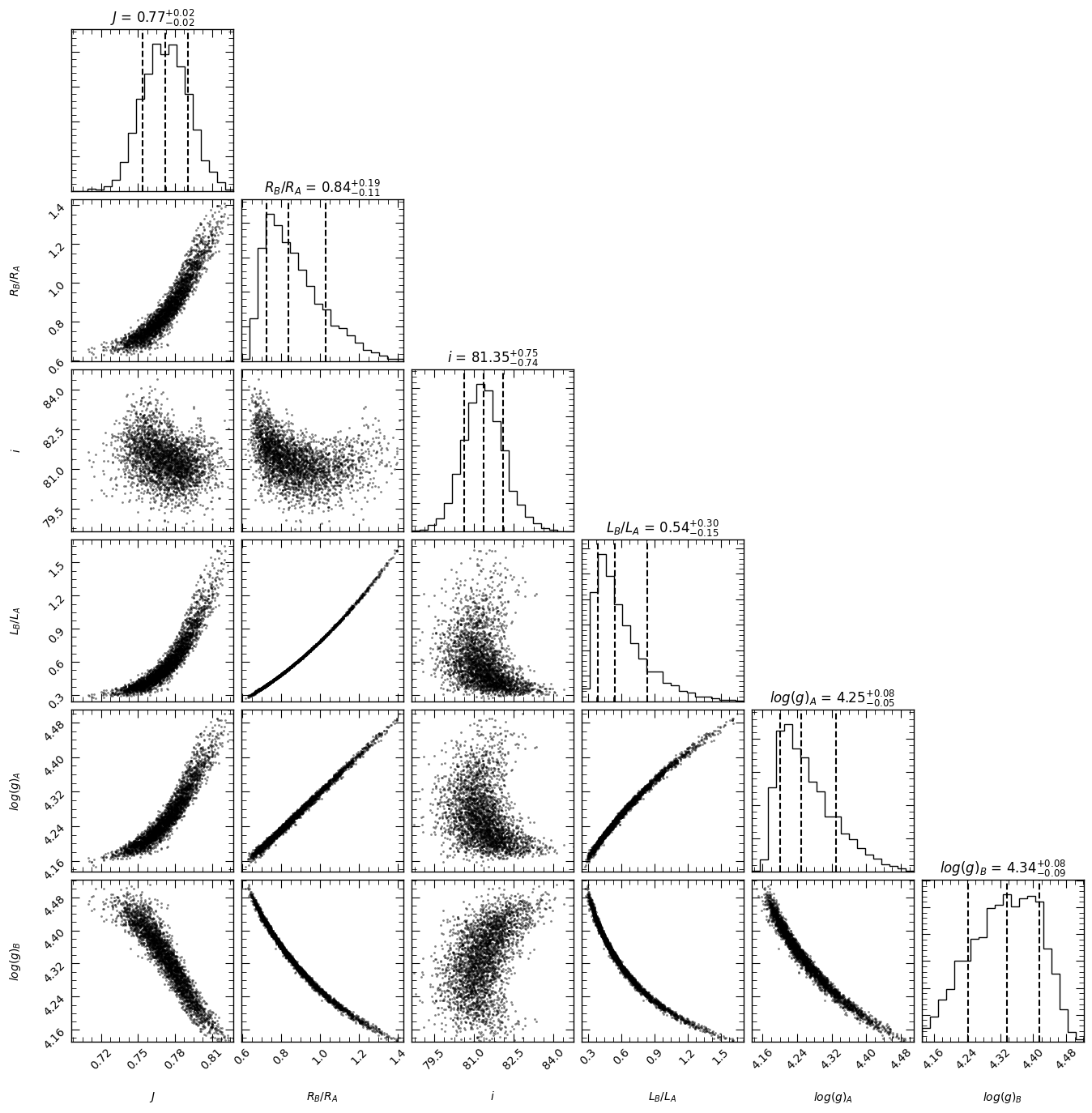}
    \caption{Corner plot for the residual-permutation simulations of TESS 45 solution. Titles show 16, 50 and 84 percentiles. }
    \label{fig:corner3}
\end{figure*}

In Figures~\ref{fig:corner}, \ref{fig:corner2}, \ref{fig:corner3} we show corner plots \citep{corner} with all residual-permutation simulation results for $J$, $R_B/R_A$, $i$, $L_B/L_A$, and $\logg_{A,B}$ for ZTF $r$ and TESS LCs.
There is strong correlation between $J$ and $R_B/R_A$ clearly visible for all datasets, which can propagate into other parameters like $L$ and $\logg$. However the most precise TESS 44 dataset is less affected.


\section{Spectral separation.}
\label{separation}

There are many spectral separation/disentangling methods available in the literature, so we tested two of them: ``shift and add'' and Fourier separation on the spectra of our SB2 system.
\par
We used the ``shift and add'' algorithm described in \citet{gonzalez2006} to extract rest-frame spectra for both components. We selected 24 spectra with $\snr\geq40$ and $\Delta\rv>180~\kms$. All these spectra were previously normalised during the binary model fitting as described in Section~\ref{sec:ind}. Radial velocities were computed using the orbital solution. Spectral components were found via iterative subtraction of the weighted sum of the Doppler-shifted spectra from the original flat baseline spectra, which were set to 0.5. The seven iterations were enough to clearly reveal both spectral components, except for the spectral edges, where solutions start to oscillate around the baseline. This approach can utilize errors of the original spectra and efficiently remove spectral artifacts, like contamination by cosmic rays.   
\par
For the Fourier separation we used {\sc FD3} code \citep{fd3}. This code can do disentangling of up to three spectral components simultaneously with the orbital fitting for inner binary system. We found out that spectral artifacts, like cosmic rays spikes, will contaminate resulting spectra, thus we used code with stacked spectra, which have been already cleaned. Blue and red parts of the spectra were analysed separately. We used 14 spectra taken in the out-of-eclipse phases with the light factors $LF=1$ for both components, thus resulting spectra should have the baseline 0.5. Derived orbital parameters are listed in Table~\ref{tab:fd3}. They are consistent with the  {\sc GLS} and {\sc JKTEBOP} orbital solutions, although $K$ and $e$ were slightly different, possibly due to usage of the stacked spectra. The resulting separated spectra showed the smooth undulations, which were removed following the steps in \cite{fd3norm}.    

\begin{table}
\begin{center}
\caption{{\sc FD3} orbital solution for stacked spectra. Unfortunately {\sc FD3} don't provide any uncertainties. }
\begin{tabular}{lcc}
\hline
Parameter & Blue & Red\\
\hline
$P$, d & 1.217784 & 1.217784\\
$t_0$, HJD d & $2458808.034$ &$2458808.034$\\
$K_A,\,\kms$ & 121.83 & 120.92\\
$K_B,\,\kms$ & 141.62 & 140.18\\
$e$      &    0.02 &   0.02 \\
$\omega\, \degr$ & 39.95 & 41.7\\
\hline
\label{tab:fd3}
\end{tabular}
\end{center}
\end{table}

\begin{figure*}
	\includegraphics[width=\textwidth]{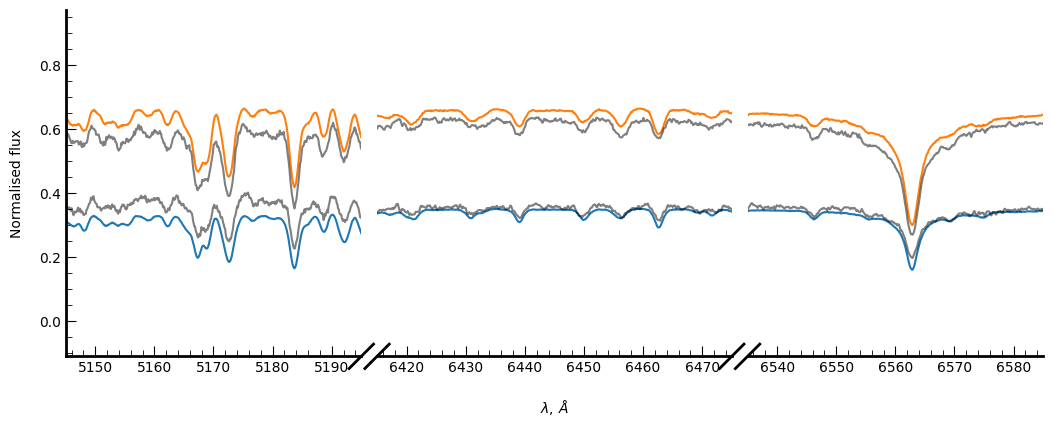}
	\includegraphics[width=\textwidth]{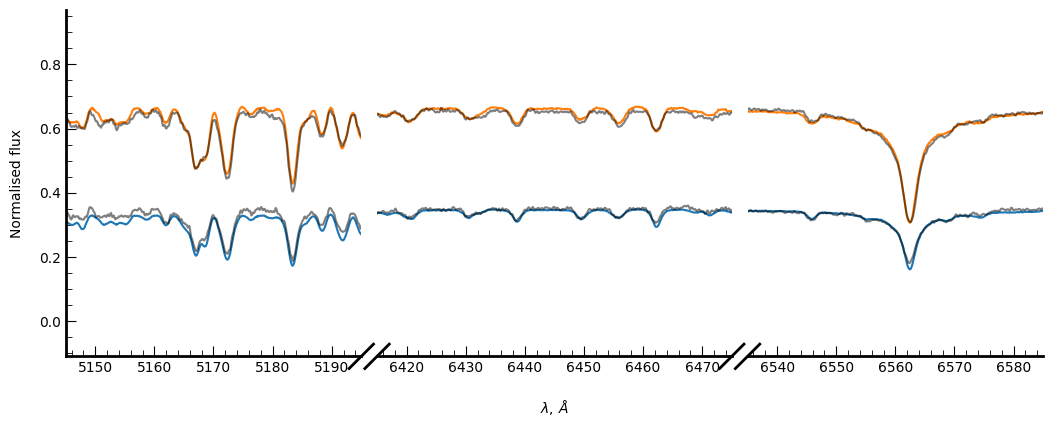}
    \caption{Comparison of the best-fit spectral models (blue and orange lines) with the separated spectra (gray lines) with offsets $+0.15,-0.15$ for primary and secondary. The top panel shows model for short epoch's spectra, fitted with the same $\feh$ assumption and ``shift and add" separation results. The bottom panel shows model for stacked spectra, fitted with the same $\feh$ assumption and Fourier separation results. Note that Fourier separation spectra are blue-shifted according to $\gamma$.}
    \label{fig:sepa}
\end{figure*}

We show the best-fit multiple-epoch spectral models with the separation results from both methods in Figure~\ref{fig:sepa}. Note, that the Fourier separation spectra are blue-shifted according to $\gamma$, since this parameter was not included in orbital solution. Generally agreement is good for {\sc FD3} results, while ``shift and add'' separation show large difference around $\ha$. Such bad performance on the broad lines is known for this method \citep{quintero2020}.    
\par
\begin{figure*}
	\includegraphics[width=\textwidth]{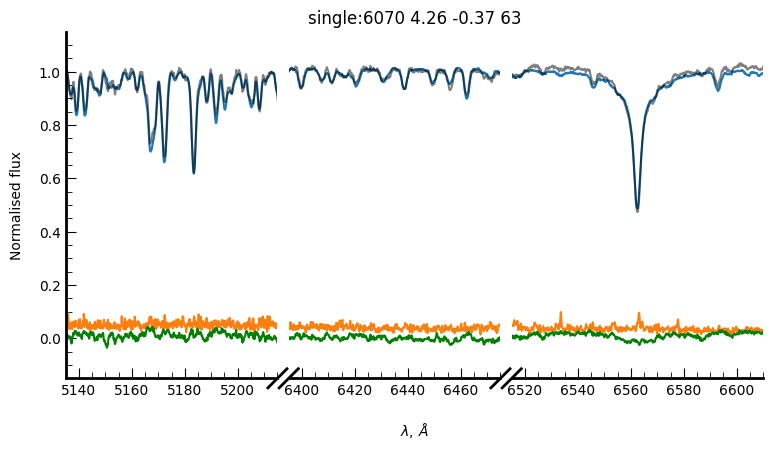}
	\includegraphics[width=\textwidth]{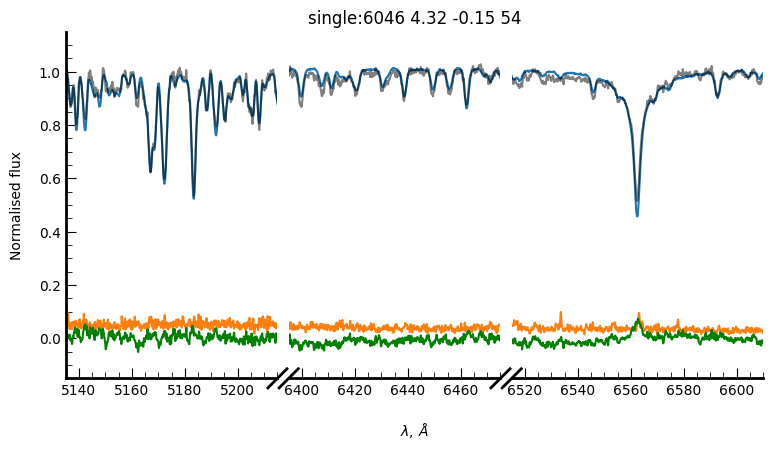}
    \caption{Separated spectra fitted by the single-star spectral model. The observed spectrum is shown as a gray line, the best fit is shown as blue line. The orange line is an error spectrum and the difference O-C is shown as a green line. The primary component is shown on the top panel, the secondary is shown on the bottom panel. }
    \label{fig:fd3}
\end{figure*}

We tried to fit the {\sc FD3} separated spectra by the single-star spectral model, but found the poor results for the secondary, not consistent with the LC, SED and multiple-spectra solutions. We used root mean square error of disentangling residuals as an error for the separated spectra, equal for both components. When we fitted for spectral parameters and light factors $LF_i$ (one for each blue and red spectral parts) we found good fits with significant difference in the metallicity $\Delta \feh>0.5$ dex between components. Also the derived light factors were not consistent for the primary and the secondary ($LF_A+LF_B\neq1$) for both blue and red parts of the spectra. We repeated fitting, fixing the light factors to values $LF_A,LF_B=0.65,\,0.35$ and keep $\logg_{A,B}=4.26,\,4.32$ (cgs) equal to values from the TESS 44 LC solution. We got a solution with ${\teff}_{A,B}=6070\pm36,\,6046\pm64$ K, $\feh_{A,B}=-0.37\pm0.03,\, -0.15\pm0.05$ dex and $\vsini_{A,B}=63\pm3,\,54\pm3~\kms$, see Figure~\ref{fig:fd3}. This result is similar to the SED temperatures and spectral parameters derived from the stacked spectra using the multiple-epoch fitting. 
The spectral separation methods require more prior information ($LF$ or orbital solution) in comparison with multiple-epochs fitting, therefore we prefer later one for analysis on LAMOST-MRS data. However for very high-quality spectra of the bright stars spectral separation can give reasonable $LF$ \citep[see][analysis of $\alpha$ Dra]{pavlovsli2022}.  

\section{Gaia DR3 Monte Carlo Multiple Star Classifier sample}

\begin{figure*}
    \includegraphics[width=\textwidth]{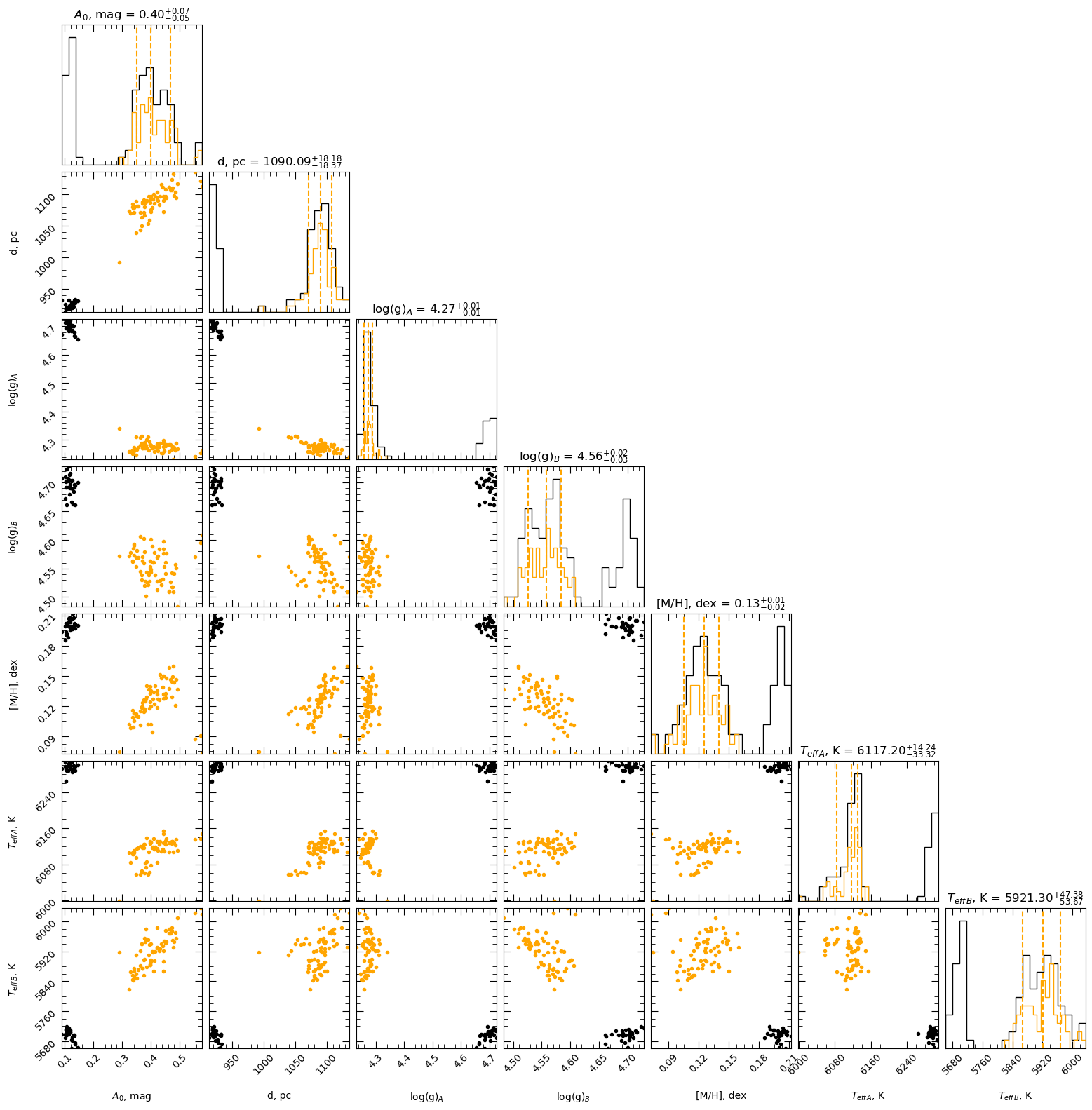}
    \caption{Corner plot for Gaia DR3 Multiple Star Classifier sample. Titles show 16, 50 and 84 percentiles for subsample with $A_0>0.2$ mag. }
    \label{fig:msc}
\end{figure*}

In Figure~\ref{fig:msc} we show corner plot \citep{corner} with Monte Carlo Multiple Star Classifier sample from \cite{gaia3}.
We select a subsample with $A_0>0.2$ mag (70 datapoints). These results have higher \texttt{log\_post}. We compute 16, 50 and 84 percentiles for them and the median values are comparable to our final parameters, however metallicity and $\logg_B$ are higher. The remaining 30 datapoints with smaller \texttt{log\_post} have $\logg_{A,B}\sim4.7$ cgs. This is probably solutions from local maximum of the posterior. 

\bsp	
\label{lastpage}
\end{document}